\documentstyle[11pt,dina4,amsfont12]{article}

\input psfig.sty

\newcommand{\p}{\partial}
\newcommand{\og}{\omega}
\newcommand{\Og}{\Omega}
\newcommand{\fl}[2]{\frac{#1}{#2}}

\newcommand{\tm}{\times}

\newcommand{\nn}{\nonumber}

\newcommand{\bt}{\beta}

\newcommand{\gm}{\gamma}

\newcommand{\tht}{\theta}

\newcommand{\vep}{\varepsilon}

\newcommand{\sg}{\sigma}

\newcommand{\btd}{\nabla}

\renewcommand{\theequation}{\arabic{section}.\arabic{equation}}
\newcommand{\be}{\begin{equation}}
\newcommand{\ee}{\end{equation}}
\newcommand{\ba}{\begin{array}}
\newcommand{\ea}{\end{array}}
\newcommand{\bea}{\begin{eqnarray}}
\newcommand{\eea}{\end{eqnarray}}
\newcommand{\beas}{\begin{eqnarray*}}
\newcommand{\eeas}{\end{eqnarray*}}

\newtheorem{remark}{Remark}[section]
\newtheorem{theorem}{Theorem}[section]
\newtheorem{lemma}{Lemma}[section]

 \newcommand{\bx}{{\bf x} }
 \newcommand{\bn}{{\bf n} }
\newtheorem{proposition}[theorem]{Proposition}

\begin{document}

\title{Efficient numerical methods for computing ground states and
dynamics of dipolar Bose-Einstein condensates}
\author{\centerline{{\it Weizhu Bao}$^{1,2,}$\thanks{Corresponding author.
Emails:\small{\,\it bao@math.nus.edu.sg(W. Bao)}, { \it
caiyongyong@nus.edu.sg(Y. Cai),}
{\,\it hanquan.wang@gmail.com(H. Wang).}}, \it{ Yongyong Cai}$^1$
and  \it{Hanquan Wang}$^{1,3}$} \\
{\footnotesize $^1$Department of Mathematics, National University of
Singapore, 117543, Singapore}\\ {\footnotesize $^2$Center for
Computational Science and Engineering, National University of
Singapore, 117543, Singapore}\\
 {\footnotesize $^4$School of Statistics
and Mathematics, Yunnan University of Finance and Economics, 
PR China} }

\date{}
\maketitle

\begin{abstract}
New efficient and accurate numerical methods are proposed to compute
ground states and  dynamics of  dipolar Bose-Einstein condensates
(BECs) described by a three-dimensional (3D) Gross-Pitaevskii
equation (GPE) with a dipolar interaction potential. Due to the high
singularity in the dipolar interaction potential, it brings
significant difficulties in mathematical analysis and numerical
simulations of dipolar BECs. In this paper, by decoupling the
two-body dipolar interaction potential into  short-range (or local)
and long-range interactions (or repulsive and attractive
interactions), the GPE for dipolar BECs is reformulated as a
Gross-Pitaevskii-Poisson type system. Based on this new mathematical
formulation, we prove rigorously existence and uniqueness as well as
nonexistence of the ground states,  and discuss the existence of
global weak solution and finite time blowup of the dynamics in
different parameter regimes of dipolar BECs. In addition,  a
backward Euler sine pseudospectral method is presented for computing
the ground states and a time-splitting sine pseudospectral  method
is proposed for computing the dynamics of dipolar BECs. Due to the
adaption of new mathematical formulation, our new numerical methods
avoid evaluating integrals with high singularity and thus they are
more efficient and accurate than those numerical methods currently
used in the literatures for solving the problem.
 Extensive numerical examples
 in 3D are reported to demonstrate the
efficiency and accuracy of our new numerical methods for computing
the ground states and dynamics of dipolar BECs.

\end{abstract}



\section{Introduction}
Since 1995, the Bose-Einstein condensation (BEC) of ultracold atomic
and molecular gases has attracted considerable interests both
theoretically and experimentally. These trapped quantum gases are
very dilute and  most of their properties are governed by the
interactions between particles in the condensate \cite{Pitaevskii}.
In the last several years, there has been a quest for realizing a
novel kind of quantum gases with the dipolar interaction, acting
between particles having a permanent magnetic or electric dipole
moment. A major breakthrough has been very recently performed at
Stuttgart University, where a BEC of ${}^{52}$Cr atoms has been
realized in experiment and it allows the experimental investigations
of the unique properties of dipolar quantum gases \cite{Griesmaier}.
In addition, recent experimental developments on cooling and
trapping of molecules \cite{Ellio}, on photoassociation \cite{Wang},
and on Feshbach resonances of binary mixtures  open much more
exciting perspectives towards a degenerate quantum gas of polar
molecules \cite{Sage}.  These success of experiments have spurred
great excitement in the atomic physics community and renewed
interests in  studying the ground states
\cite{Santos,Yi,Goral,Goral1,Jiang,Ronen} and dynamics
\cite{Lahaye1,Parker, Pedri,Yi1} of dipolar BECs.

At temperature $T$ much smaller than the critical temperature $T_c$,
a dipolar BEC is well described by the macroscopic wave function
$\psi=\psi(\bx,t)$ whose evolution is governed by the
three-dimensional (3D) Gross-Pitaevskii equation (GPE)
\cite{Yi,Santos} \be \label{ngpe} i\hbar \p_t
\psi(\bx,t)=\left[-\fl{\hbar^2}{2m}\nabla^2+V(\bx)+U_0|\psi|^2+
\left(V_{\rm dip}\ast |\psi|^2\right)\right]\psi, \quad \bx\in{\Bbb
R}^3, \ t>0, \ee where $t$ is time, $\bx=(x,y,z)^T\in {\Bbb R^3}$ is
the Cartesian coordinates, $\hbar$ is the Planck constant, $m$ is
the mass of a dipolar particle and $V(\bx)$ is an external trapping
potential. When a harmonic trap potential is considered,
$V(\bx)=\fl{m}{2}(\omega_{x}^2x^2+ \omega_y^2y^2+\omega_{z}^2 z^2)$
with $\omega_x$, $\omega_y$ and $\omega_z$ being the trap
frequencies in $x$-, $y$- and $z$-directions, respectively.
$U_0=\frac{4\pi \hbar^2 a_s}{m}$ describes local (or short-range)
interaction between dipoles in the condensate with $a_s$ the
$s$-wave scattering length (positive for repulsive interaction and
negative for attractive interaction). The long-range dipolar
interaction potential between two dipoles is given by
\be\label{kel0} V_{\rm dip}(\bx)= \frac{\mu_0\mu_{\rm
dip}^2}{4\pi}\,\fl{1-3(\bx\cdot \bf
n)^2/|\bx|^2}{|\bx|^3}=\frac{\mu_0\mu_{\rm
dip}^2}{4\pi}\,\fl{1-3\cos^2(\theta)}{|\bx|^3}, \qquad \bx\in{\Bbb
R}^3,\ee  where $\mu_0$ is the vacuum magnetic permeability,
$\mu_{\rm dip}$ is permanent magnetic dipole moment (e.g. $\mu_{\rm
dip}=6\mu_{_B}$ for $^{52}$C$_{\rm r}$ with $\mu_{_B}$ being the
Bohr magneton), ${\bf n}=(n_1,n_2,n_3)^T\in {\Bbb R}^3$ is the
dipole axis (or dipole moment) which is a given unit vector, i.e.
$|{\bf n}|=\sqrt{n_1^2+n_2^2+n_3^3}=1$, and $\theta$ is the angle
between the dipole axis ${\bf n}$ and the vector $\bx$. The wave
function is normalized according to \be\label{norm00}
\|\psi\|^2:=\int_{{\Bbb R}^d} |\psi(\bx,t)|^2\;d\bx =N,\ee where $N$
is the total number of dipolar particles in the dipolar BEC.

By introducing the dimensionless variables, $t\to \frac{t}{\og_{0}}$
with $\og_0=\min\{\og_x,\og_y,\og_z\}$,  $\bx \to a_0\bx$ with $
a_0=\sqrt{\fl{\hbar}{m\og_{0}}}$, $\psi\to \frac{\sqrt{N}
\psi}{a_0^{3/2}}$,
 we obtain the dimensionless GPE  in 3D from (\ref{ngpe})
 as \cite{Yi,Yi2,Pitaevskii,Bao_Jaksch_Markowich}:
 \be \label{ngpe1} i\p_t \psi(\bx,t)=\left[-\fl{1}{2}\nabla^2+V(\bx)+\beta
|\psi|^2+\lambda \left(U_{\rm dip}\ast|\psi|^2\right)\right]\psi,
\qquad \bx\in{\Bbb R}^3, \quad t>0,\ee where $\beta
=\frac{NU_0}{\hbar\og_0 a_0^3}= \fl{4\pi a_sN}{a_0}$, $\lambda
=\fl{mN\mu_0\mu_{\rm dip}^2}{3\hbar^2 a_0}$,
$V(\bx)=\fl{1}{2}(\gamma_{x}^2x^2+ \gamma_y^2y^2+\gamma_{z}^2 z^2)$
is the dimensionless harmonic trapping potential with
$\gm_x=\frac{\og_x}{\og_0}$, $\gm_y=\frac{\og_y}{\og_0}$ and
$\gm_z=\frac{\og_z}{\og_0}$, and the dimensionless long-range
dipolar interaction potential $U_{\rm dip}(\bx)$ is given as
\be\label{kel} U_{\rm dip}(\bx)= \frac{3}{4\pi}\,\fl{1-3(\bx\cdot
\bf n)^2/|\bx|^2}{|\bx|^3}=\frac{3}{4\pi}\,
\fl{1-3\cos^2(\theta)}{|\bx|^3}, \qquad \bx\in{\Bbb R}^3.\ee From
now on, we will treat $\beta$ and $\lambda$ as two dimensionless
real parameters. We understand that it may not physical meaningful
when $\lambda<0$ for modeling dipolar BEC. However, it is an
interesting problem to consider the case when $\lambda<0$ at least
in mathematics and it may make sense for modeling other physical
system. In fact, the above nondimensionlization is obtained by
adopting a unit system where the units for length, time and energy
are given by $a_0$, $1/\og_0$ and $\hbar \og_0$, respectively. Two
important invariants of (\ref{ngpe1}) are the {\sl mass} (or
normalization) of the wave function
 \be\label{norm}
N(\psi(\cdot,t)):=\|\psi(\cdot,t)\|^2= \int_{{\Bbb R}^3}
|\psi(\bx,t)|^2\;d \bx\equiv \int_{{\Bbb R}^3} |\psi(\bx,0)|^2\;d
\bx=1, \qquad t\ge0, \ee and the {\sl energy} per particle \bea
 E(\psi(\cdot,t))&:=&\int_{{\Bbb R}^3}\left[
\frac{1}{2}|\nabla\psi |^2 +V(\bx) |\psi|^2 +\fl{\beta}{2}|\psi|^4 +
\fl{\lambda}{2}\left(U_{\rm dip}\ast|\psi|^2\right) |\psi|^2
\right]d \bx\nn\\
&\equiv& E(\psi(\cdot,0)), \qquad t\ge0. \label{energy} \eea

To find the stationary states including ground and excited states of
a dipolar BEC, we take the ansatz \be \label{stat}
\psi(\bx,t)=e^{-i\mu t}\phi (\bx), \qquad \bx\in{\Bbb R}^3, \quad
t\ge0,\ee where $\mu\in{\Bbb R}$ is the chemical potential and
$\phi:=\phi(\bx)$ is a time-independent function. Plugging
(\ref{stat}) into (\ref{ngpe1}), we get the time-independent GPE (or
a nonlinear eigenvalue problem) \bea
 \label{gpe22dstat} \mu\, \phi(\bx)=\left[-\fl{1}{2 }\btd^2
+V(\bx)+\beta|\phi|^2+\lambda\left(U_{\rm
dip}\ast|\phi|^2\right)\right] \phi(\bx),\qquad \bx\in{\Bbb R}^3,
\eea under the constraint \be\label{const} \|\phi\|^2:=\int_{{\Bbb
R}^3}|\phi(\bx)|^2\;d\bx=1.\ee
 The ground state of a dipolar BEC
is usually defined as
the minimizer of the following nonconvex minimization problem:\\
Find $\phi_g \in S $ and $\mu^g\in {\Bbb R}$   such that \be
\label{groundstate} E^g:=E(\phi_g)=\min_{\phi\in S}\ E (\phi),\qquad
\mu^g:=\mu(\phi_g), \ee where the nonconvex set $S$ is defined as
\be S:=\left\{\phi \ |\  \|\phi\|^2=1, \, E(\phi)<\infty \right\}
\ee and the chemical potential (or eigenvalue of (\ref{gpe22dstat}))
is defined as \bea
 \mu(\phi)&:=&\int_{{\Bbb R}^3}\left[
\frac{1}{2}|\nabla\phi |^2 +V(\bx) |\phi|^2 +\beta|\phi|^4 +
\lambda\left(U_{\rm dip}\ast|\phi|^2\right) |\phi|^2
\right]d \bx\nn\\
&\equiv& E(\phi)+\frac{1}{2}\int_{{\Bbb R}^3}\left[ \beta|\phi|^4 +
\lambda\left(U_{\rm dip}\ast|\phi|^2\right) |\phi|^2 \right]d \bx.
\label{chem00} \eea In fact, the nonlinear eigenvalue problem
(\ref{gpe22dstat}) under the constraint (\ref{const}) can be viewed
as the Euler-Lagrangian equation of the nonconvex minimization
problem (\ref{groundstate}). Any eigenfunction of the nonlinear
eigenvalue problem (\ref{gpe22dstat}) under the constraint
(\ref{const}) whose energy is larger than that of the ground state
is usually called as an excited state in the physics literatures.

The theoretical study of dipolar BECs including ground states and
dynamics as well as quantized vortices has been carried out in
recent years based on the GPE (\ref{ngpe}). For the study in
physics, we refer to
\cite{Eberlein,Giovanazzi,Klawunn,Recati,Abad,Glaum,Klawunn,Nath,Odell0,
Wilson,Wilson1,Yi2,Zhang} and references therein. For the study in
mathematics, existence and uniqueness as well as the possible
blow-up of solutions were studied in \cite{Carles}, and existence of
solitary waves was proven in \cite{Ant}. In most of the numerical
methods used in the literatures for theoretically and/or numerically
studying the ground states and dynamics of dipolar BECs, the way to
deal with the convolution in (\ref{ngpe1}) is usually to use the
Fourier transform \cite{Lahaye1,Goral,Ronen,Xiong,Blakie,Tick,Yi3}.
However, due to the high singularity in the dipolar interaction
potential (\ref{kel}), there are two drawbacks in these numerical
methods: (i) the Fourier transforms of the dipolar interaction
potential (\ref{kel}) and the density function $|\psi|^2$ are
usually carried out in the continuous level on the whole space
${\Bbb R}^3$ (see (\ref{four11}) for details) and in the discrete
level on a bounded computational domain $\Og$, respectively, and due
to this mismatch, there is a locking phenomena in practical
computation as observed in \cite{Ronen}; (ii) the second term in the
Fourier transform of the dipolar interaction potential is
$\frac{0}{0}$-type for $0$-mode, i.e when $\xi=0$ (see
(\ref{four11}) for details), and it is artificially omitted when
$\xi=0$ in practical computation
\cite{Ronen,Goral1,ODell,Yi1,Yi2,Xiong,Blakie} thus this may cause
some numerical problems too. The main aim of this paper is to
propose new numerical methods for computing ground states and
dynamics of dipolar BECs which can avoid the above two drawbacks and
thus they are more accurate than those currently used in the
literatures. The key step is to decouple the dipolar interaction
potential into a short-range and a long-range interaction (see
(\ref{Udip0}) for details) and thus we can reformulate the GPE
(\ref{ngpe1}) into a Gross-Pitaevskii-Poisson type system. In
addition, based on the new mathematical formulation, we can prove
existence and uniqueness as well as nonexistence of the ground
states and discuss mathematically the dynamical properties of
dipolar BECs in different parameter regimes.

The paper is organized as follows. In section 2, we reformulate the
GPE for a dipolar BEC into a Gross-Pitaevskii-Poisson type system
and study analytically the ground states and dynamics of dipolar
BECs. In section 3, a backward Euler sine pseudospectral method is
proposed for computing ground states of dipolar BECs; and in section
4, a time-splitting sine pseudospectral (TSSP) method is presented
for computing the dynamics. Extensive numerical results are reported
in section 5 to demonstrate the efficiency and accuracy of our new
numerical methods.  Finally, some conclusions are drawn in section
6. Throughout this paper, we adapt the standard Sobolev spaces and
their corresponding norms.

\section{Analytical results for ground sates and dynamics}
\setcounter{equation}{0}

Let $r=|\bx|=\sqrt{x^2+y^2+z^2}$ and denote \be\label{pbn0}
\p_\bn={\bf n}\cdot \nabla=n_1\p_x+n_2\p_y+n_3\p_z, \qquad
\p_{\bn\bn}=\p_\bn(\p_\bn).\ee Using the equality (see \cite{Parker}
and a mathematical proof in the Appendix)  \be \label{decop1}
\fl{1}{r^3}\left(1-\frac{3(\bx\cdot {\bf n})^2}{r^2}\right) =
-\fl{4\pi}{3} \delta (\bx)-\p_{\bn\bn}\left(
\frac{1}{r}\right),\qquad \bx\in {\Bbb R}^3, \label{formula}\ee with
$\delta(\bx)$ being the Dirac distribution function and  introducing
a new function \be \label{varp0} \varphi(\bx,t):=\left(\frac{1}{4\pi
|\bx|}\right) \ast |\psi(\cdot,t)|^2 = \fl{1}{4\pi} \int_{{\Bbb
R}^3} \fl{1}{|\bx-\bx'|}|\psi(\bx',t)|^2\;d\bx', \qquad \bx\in{\Bbb
R}^3, \quad t\ge0,\ee we obtain \be\label{Udip0} U_{\rm
dip}\ast|\psi(\cdot,t)|^2 = -|\psi(\bx,t)|^2-3\p_{\bn\bn}
\left(\varphi (\bx, t)\right), \qquad \bx\in {\Bbb R}^3, \quad
t\ge0. \label{integral}\ee In fact, the above equality decouples the
dipolar interaction potential into a short-range and a long-range
interaction which correspond to the first and second terms in the
right hand side of (\ref{Udip0}), respectively. In fact, from
(\ref{pbn0})-(\ref{Udip0}), it is straightforward to get the Fourier
transform of $U_{\rm dip}(\bx)$ as \be\label{four11}
\widehat{(U_{\rm dip})}(\xi)=-1+\frac{3\left(\bn\cdot
\xi\right)^2}{|\xi|^2}, \qquad \xi\in {\Bbb R}^3. \ee Plugging
(\ref{Udip0}) into (\ref{ngpe1}) and noticing (\ref{varp0}), we can
reformulate the GPE (\ref{ngpe1}) into  a Gross-Pitaevskii-Poisson
type system \bea \label{gpe} &&i \p_t
\psi(\bx,t)=\left[-\fl{1}{2}\nabla^2+V(\bx)+(\beta-\lambda)
|\psi(\bx,t)|^2-3\lambda \p_{\bn\bn} \varphi(\bx,t)
\right]\psi(\bx,t),  \\
\label{poisson}&&\qquad \nabla^2 \varphi(\bx,t) =
-|\psi(\bx,t)|^2,\qquad
\lim\limits_{|\bx|\to\infty}\varphi(\bx,t)=0\qquad \bx\in{\Bbb R}^3,
\quad t>0.
 \eea
Note that the far-field condition in (\ref{poisson}) makes the
Poisson equation uniquely solvable. Using (\ref{poisson}) and
integration by parts, we can reformulate the energy functional
$E(\cdot)$ in (\ref{energy}) as \be\label{newener}
E(\psi)=\int_{\Bbb R^3 }\left[\frac 12|\nabla
\psi|^2+V(\bx)|\psi|^2+\frac{1}{2}(\beta-\lambda
)|\psi|^4+\frac{3\lambda}{2}\left|\p_{\bf n}\nabla
\varphi\right|^2\right]\,d\bx\,,\ee where $\varphi$ is defined
through (\ref{poisson}). This immediately shows that the decoupled
short-range and long-range interactions of the dipolar interaction
potential are attractive and repulsive, respectively,  when
$\lambda>0$; and are repulsive and attractive, respectively, when
$\lambda<0$. Similarly, the nonlinear eigenvalue problem
(\ref{gpe22dstat}) can be reformulated as \bea
 \label{gpe22dstat1} &&\mu\, \phi(\bx)=\left[-\fl{1}{2 }\btd^2
+V(\bx)+\left(\beta-\lambda\right)|\phi|^2-3\lambda \p_{\bn\bn}
\varphi(\bx)\right] \phi(\bx), \\
&&\qquad  \nabla^2 \varphi(\bx) =-|\phi(\bx)|^2, \quad \bx\in{\Bbb
R}^3,\qquad
\lim\limits_{|\bx|\to\infty}\varphi(\bx)=0.\label{poi101} \eea

\subsection{Existence and uniqueness for ground states}
Under the  new formulation for the energy functional $E(\cdot)$ in
(\ref{newener}), we have

\begin{lemma}\label{lem1} For the energy  $E(\cdot)$ in (\ref{newener}), we have

(i) For any $\phi\in S$, denote $\rho(\bx)=|\phi(\bx)|^2$ for
$\bx\in{\Bbb R}^3$, then we have \be E(\phi)\geq
E(|\phi|)=E\left(\sqrt{\rho}\right), \qquad \forall \phi\in S,\ee so
the minimizer $\phi_g$ of (\ref{groundstate}) is of the form
$e^{i\theta_0}|\phi_g|$ for some constant $\theta_0\in \Bbb R$.

(ii) When $\beta\ge0$ and $-\frac12\beta\leq \lambda\leq \beta$, the
energy $E(\sqrt{\rho})$ is strictly convex in $\rho$.
\end{lemma}

\noindent {\bf Proof:} For any $\phi\in S$, denote $\rho=|\phi|^2$
and consider the Poisson equation \be
\label{poi11}\nabla^2\varphi(\bx)=-|\phi(\bx)|^2:=-\rho(\bx), \quad
\bx\in{\Bbb R}^3, \qquad
\lim\limits_{|\bx|\to\infty}\varphi(\bx)=0.\ee Noticing (\ref{pbn0})
with $|\bn|=1$,  we have the estimate \be \label{pbnf1}\|\p_{\bf
n}\nabla \varphi\|_2\leq\|D^2\varphi\|_2=\|\nabla^2\varphi\|_2=
\|\rho\|_2=\|\phi\|_4^2,\qquad \hbox{with} \quad D^2=\nabla\nabla.
\ee

(i) Write $\phi(\bx)=e^{i\theta(\bx)}|\phi(\bx)|$, noticing
(\ref{newener}) with $\psi=\phi$ and (\ref{poi11}), we get
\begin{eqnarray}
E(\phi)&=&\int_{\Bbb R^3}\left[|\,\nabla
|\phi|\,|^2+|\phi|^2|\nabla\theta(\bx)|^2+V(\bx)|\phi|^2+\frac{1}{2}(\beta-\lambda
)|\phi|^4+\frac{3\lambda}{2}|\p_{\bf n}\nabla
\varphi|^2\right]\,d\bx\nn\\
&\geq&\int_{\Bbb R^3}\left[|\,\nabla
|\phi|\,|^2+V(\bx)|\phi|^2+\frac{1}{2}(\beta-\lambda
)|\phi|^4+\frac{3\lambda}{2}|\p_{\bf n}\nabla
\varphi|^2\right]\,d\bx\nn\\
&=&E(|\phi|)=E\left(\sqrt{\rho}\right), \qquad \forall \phi\in S,
\end{eqnarray}
and the equality holds iff $\nabla\theta(\bx)=0$ for $\bx\in {\Bbb
R}^3$, which means $\theta(\bx)\equiv \tht_0$ is a constant.

(ii) From (\ref{newener}) with $\psi=\phi$ and noticing
(\ref{poi11}), we can split the energy $E\left(\sqrt{\rho}\right)$
into two parts, i.e.
\begin{eqnarray}E(\sqrt{\rho})
=E_1(\sqrt{\rho})+E_2(\sqrt{\rho}), \eea where \bea\label{e11}
&&E_1(\sqrt{\rho})=\int_{\Bbb R^3}\left[|\nabla
\sqrt{\rho}|^2+V(\bx)\rho\right]d\bx, \\ \label{e12}
&&E_2(\sqrt{\rho})=\int_{\Bbb R^3}\left[\frac{1}{2}(\beta-\lambda
)|\rho|^2+\frac{3\lambda}{2}|\p_{\bf n}\nabla
\varphi|^2\right]\,d\bx. \label{e1e2} \eea As shown in \cite{Lie},
$E_1\left(\sqrt{\rho}\right)$ is convex (strictly) in $\rho$. Thus
we need only prove $E_2\left(\sqrt{\rho}\right)$ is convex too. In
order to do so, consider $\sqrt{\rho_1}\in S$, $\sqrt{\rho_2}\in S$,
and let $\varphi_1$ and $\varphi_2$ be the solutions of the Poisson
equation (\ref{poi11}) with $\rho=\rho_1$ and $\rho=\rho_2$,
respectively. For any $\alpha\in[0,1]$, we have
$\sqrt{\alpha\rho_1+(1-\alpha)\rho_2}\in S$, and
\begin{eqnarray}
\lefteqn{\alpha E_2(\sqrt{\rho_1})+(1-\alpha)E_2(\sqrt{\rho_2})-E_2
\left(\sqrt{\alpha\rho_1+(1-\alpha)\rho_2}\right)\nn}\\[2mm]
&=&\alpha(1-\alpha)\int_{\Bbb R^3}\left[\frac{1}{2}(\beta-\lambda
)(\rho_1-\rho_2)^2+\frac{3\lambda}{2}|\p_{\bf n}\nabla
(\varphi_1-\varphi_2)|^2\right]\,d\bx,\label{poi00}
\end{eqnarray}
which immediately implies that $E_2(\sqrt{\rho})$ is convex if
$\beta\ge0$ and $0\leq\lambda\leq \beta$.  If $\beta\ge0$ and
$-\frac{1}{2}\beta\leq \lambda<0$, noticing that
$\alpha\varphi_1+(1-\alpha)\varphi_2$ is the solution of the Poisson
equation (\ref{poi11}) with $\rho=\alpha\rho_1+(1-\alpha)\rho_2$,
combining (\ref{pbnf1}) with $\varphi=\varphi_1-\varphi_2$ and
(\ref{poi00}), we obtain $E_2(\sqrt{\rho})$ is convex again.
Combining all the results above together, the conclusion follows.
$\Box$

Now, we are able to prove the existence and uniqueness as well as
nonexistence results for the ground state of a dipolar BEC in
different parameter regimes.

\begin{theorem} Assume $V(\bx)\ge0$ for $\bx\in{\Bbb R}^3$ and
$\lim\limits_{|\bx|\to\infty}V(\bx)=\infty$ (i.e., confining
potential), then we have:

(i) If $\beta\ge0$ and $-\frac{1}{2}\beta\leq \lambda\leq \beta$,
there exists a ground state $\phi_g\in S$, and the positive ground
state $|\phi_g|$ is unique. Moreover, $\phi_g=e^{i\theta_0}|\phi_g|$
for some constant $\theta_0\in\Bbb R$.

(ii) If $\beta<0$, or $\beta\ge0$ and $\lambda<-\frac{1}{2}\beta$ or
$\lambda>\beta$, there exists no ground state, i.e.,
$\inf\limits_{\phi\in S}E(\phi)=-\infty$.
\end{theorem}

\noindent {\bf Proof:} (i) Assume $\beta\ge0$ and
$-\frac{1}{2}\beta\leq \lambda\leq \beta$, we first show $E(\phi)$
is nonnegative in $S$, i.e. \be\label{Ephi} E(\phi)=\int_{\Bbb
R^3}\left[|\nabla \phi|^2+V(\bx)|\phi|^2+\frac{1}{2}(\beta-\lambda
)|\phi|^4+\frac{3\lambda}{2}|\p_{\bf n}\nabla
\varphi|^2\right]\,d\bx\ge0, \qquad \forall \phi\in S.\ee In fact,
when $\beta\ge0$ and $0\leq\lambda\leq \beta$, noticing
(\ref{newener}) with $\psi=\phi$, it is obvious that (\ref{Ephi}) is
valid. When $\beta\ge0$ and  $-\frac 12\beta\leq \lambda<0$,
combining  (\ref{newener}) with $\psi=\phi$, (\ref{poi11}) and
(\ref{pbnf1}), we obtain (\ref{Ephi}) again as \bea
E(\phi)&\ge&\int_{\Bbb R^3}\left[|\nabla
\phi|^2+V(\bx)|\phi|^2+\frac{1}{2}(\beta-\lambda)|\phi|^4
+\frac{3\lambda}{2}|\phi|^4\right]\,d\bx\nn\\
&=&\int_{\Bbb R^3}\left[|\nabla
\phi|^2+V(\bx)|\phi|^2+\frac{1}{2}\left(\beta+2\lambda\right)|\phi|^4
\right]\,d\bx\ge0. \eea Now, let $\{\phi^n\}_{n=0}^\infty\subset S$
be a minimizing sequence of the minimization problem
(\ref{groundstate}). Then there exists a constant $C$ such that \be
\|\nabla\phi^n\|_2\le C, \qquad \|\phi^n\|_4\le C, \qquad \int_{\Bbb
R^3} V(\bx)|\phi^n(\bx)|^2d\bx\le C, \qquad n\ge0.\ee Therefore
$\phi^n$ belongs to a weakly compact set in $L^4$, $H^1=\{\phi\ |\
\|\phi\|_2+\|\nabla \phi\|_2<\infty\}$, and $L^2_V=\{\phi\ |\
\int_{{\Bbb R}^3} V(\bx) |\phi(\bx)|^2\; d\bx<\infty\}$ with a
weighted $L^2$-norm given by $\|\phi\|_V=[\int_{{\Bbb
R}^3}|\phi(\bx)|^2V(\bx)d\bx]^{1/2}$. Thus, there exists a
$\phi^\infty\in H^1\bigcap L^2_V\bigcap L^4$ and a subsequence
(which we denote as the original sequence for simplicity), such that
\be \label{conveg0}
 \phi^n\rightharpoonup\phi^\infty,\quad  \mbox{in } L^2\cap L^4\cap
 L^2_V,\qquad\quad
\nabla \phi^n\rightharpoonup\nabla\phi^\infty,\quad \mbox{in } L^2.
\ee Also, we can suppose that $\phi^n$ is nonnegative, since we can
replace them with $|\phi^n|$ , which also minimize the functional
$E$.  Similar as in \cite{Lie}, we can obtain $\|\phi^\infty\|_2=1$
due to the confining property of the potential $V(\bx)$. So,
$\phi^\infty\in S$. Moreover, the $L^2$-norm convergence of $\phi^n$
and weak convergence in (\ref{conveg0}) would imply the strong
convergence $\phi^n\to\phi^\infty\in L^2$. Thus, employing
H\"{o}lder inequality and Sobolev inequality, we obtain
\begin{eqnarray}\lefteqn{\|(\phi^n)^2-(\phi^\infty)^2\|_2\leq
C_1\|\phi^n-\phi^\infty\|_2^{1/2}(\|\phi^n\|_6^{1/2}+\|\phi^\infty\|_6^{1/2})\nn}
\\[2mm]
&\leq&
C_2(\|\nabla\phi^n\|_2^{1/2}+\|\nabla\phi^\infty\|_2^{1/2})\|\phi^n-\phi^\infty\|_2\to
0,\qquad n\to \infty,\end{eqnarray} which shows
$\rho^n=(\phi^n)^2\to \rho^\infty=(\phi^\infty)^2 \in L^2$. Since
$E_2(\sqrt{\rho})$ in (\ref{e12}) is convex and  lower
semi-continuous in $\rho$, thus
$E_2(\phi^\infty)\leq\lim\limits_{n\to\infty}E_2(\phi^n)$. For $E_1$
in  (\ref{e11}),
$E_1(\phi^\infty)\leq\lim\limits_{n\to\infty}E_1(\phi^n)$ because of
the lower semi-continuity of the $H^1$- and $L^2_V$-norm. Combining
the results together, we know
$E(\phi^\infty)\leq\lim\limits_{n\to\infty}E(\phi^n)$, which proves
that $\phi^\infty$ is indeed a minimizer of the minimization problem
(\ref{groundstate}). The uniqueness follows from the strictly
convexity of $E(\sqrt{\rho})$ as shown in Lemma \ref{lem1}.

(ii) Assume $\beta<0$, or $\beta\ge0$ and
$\lambda<-\frac{1}{2}\beta$ or $\lambda>\beta$. Without loss of
generality,  we assume ${\bf n}=(0,0,1)^T$ and choose the function
\be\label{phiv11}\phi_{\vep_1,\vep_2}(\bx)=\frac{1}{(2\pi\vep_1)^{1/2}}
\cdot\frac{1}{(2\pi\vep_2)^{1/4}}\exp\left(-\frac{x^2+y^2}
{2\vep_1}\right)\exp\left(-\frac{z^2}{2\vep_2}\right), \qquad
\bx\in{\Bbb R}^3, \ee with $\vep_1$ and $\vep_2$ two small positive
parameters (in fact, for general ${\bf n}\in {\Bbb R}^3$ satisfies
$|\bn|=1$, we can always choose $0\ne {\bf n}_1\in{\Bbb R}^3$ and
$0\ne {\bf n}_2\in{\Bbb R}^3$ such that  $\{{\bf n}_1,\, {\bf
n}_2,\,\bn\}$ forms an orthonormal basis of $\Bbb R^3$ and do the
change of variables $\bx=(x,y,z)^T$ to ${\bf y}=(\bx\cdot{\bf
n}_1,\,\bx\cdot{\bf n}_2,\,\bx\cdot{\bf n})^T$ on the right hand
side of (\ref{newener}), the following computation is still valid).
Taking the standard Fourier transform at both sides of the Poisson
equation \be -\nabla^2
\varphi_{\vep_1,\vep_2}(\bx)=|\phi_{\vep_1,\vep_2}(\bx)|^2=\rho_{\vep_1,\vep_2}(\bx),
\quad \bx\in{\Bbb R}^3, \qquad
\lim\limits_{|\bx|\to\infty}\varphi_{\vep_1,\vep_2}(\bx)=0,\ee we
get \be
|\xi|^2\widehat{\varphi_{\vep_1,\vep_2}}(\xi)=\widehat{\rho_{\vep_1,\vep_2}}(\xi),
\qquad \xi\in{\Bbb R}^3. \ee Using the Plancherel formula and
changing of variables, we obtain
\begin{eqnarray} \label{pbfn3}
\|\p_{\bf n}\nabla
\varphi_{\vep_1,\vep_2}\|_2^2&=&\frac{1}{(2\pi)^3}\|({\bf n}\cdot {
\xi})\widehat{\varphi_{\vep_1,\vep_2}}(\xi)\|_2^2
=\frac{1}{(2\pi)^3}\int_{\Bbb
R^3}\frac{|\xi_3|^2}{|\xi|^2}\left(\widehat{\rho_{\vep_1,\vep_2}}(\xi)\right)^2d\xi\nn\\
&=&\frac{1}{(2\pi)^3\vep_1\sqrt{\vep_2}}\int_{\Bbb
R^3}\frac{|\xi_3|^2}{(|\xi_1|^2+|\xi_2|^2)\cdot\frac{\vep_2}
{\vep_1}+|\xi_3|^2}\left(\widehat{\rho_{1,1}}(\xi)\right)^2\,d\xi,
\quad \vep_1,\vep_2>0.\qquad\quad
\end{eqnarray}
By the dominated convergence theorem, we get \bea
 \|\p_{\bf n}\nabla
\varphi_{\vep_1,\vep_2}\|_2^2\to\left\{\ba{ll} 0,
&\vep_2/\vep_1\to+\infty, \\
\frac{1}{(2\pi)^3\vep_1\sqrt{\vep_2}}\displaystyle\int_{\Bbb
R^3}\left(\widehat{\rho_{1,1}}(\xi)\right)^2\,d\xi
=\|\rho_{\vep_1,\vep_2}\|_2^2=\|\phi_{\vep_1,\vep_2}\|_4^4,
&\vep_2/\vep_1\to0^+. \ea\right.\qquad \eea When fixed
$\vep_1\sqrt{\vep_2}$, the last integral in (\ref{pbfn3}) is
continuous in $\vep_2/\vep_1>0$. Thus, for any $\alpha\in(0,1)$, by
adjusting $\vep_2/\vep_1:=C_\alpha>0$, we could have $\|\p_{\bf
n}\nabla
\varphi_{\vep_1,\vep_2}\|_2^2=\alpha\|\phi_{\vep_1,\vep_2}\|_4^4$.
Substituting (\ref{phiv11}) into (\ref{e11}) and (\ref{e12}) with
$\sqrt{\rho}=\phi_{\vep_1,\vep_2}$ under fixed $\vep_2/\vep_1>0$, we
get \bea E_1(\phi_{\vep_1,\vep_2})&=&\int_{\Bbb R^3}\left[|\nabla
\phi_{\vep_1,\vep_2}|^2+V(\bx)|\phi_{\vep_1,\vep_2}|^2\right]\,
d\bx=\fl{C_1}{\vep_1}+
\fl{C_2}{\vep_2}+{\mathcal{O}}(1), \qquad \\
E_2(\phi_{\vep_1,\vep_2})&=&\frac12\int_{\Bbb
R^3}(\beta-\lambda+3\alpha\lambda)
)|\phi_{\vep_1,\vep_2}|^4\,d\bx=\frac{\beta-\lambda+3\alpha
\lambda}{2}\cdot\frac{C_3}{\vep_1\sqrt{\vep_2}}, \eea with some
constants $C_1$, $C_2$, $C_3>0$ independent of $\vep_1$ and
$\vep_2$. Thus, if $\beta<0$, choose $\alpha=1/3$; if $\beta\ge0$
and $\lambda< -\frac12\beta$, choose
$1/3-\frac{\beta}{3\lambda}<\alpha<1$; and if $\beta\ge0$ and
$\lambda>\beta$, choose
$0<\alpha<\frac{1}{3}\left(1-\frac{\beta}{\lambda}\right)$; as
$\vep_1$, $\vep_2\to 0^+$, we can get $\inf\limits_{\phi\in
S}E(\phi)=\lim\limits_{\vep_1,\vep_2\to 0^+}
E_1(\phi_{\vep_1,\vep_2}) +E_2(\phi_{\vep_1,\vep_2})=-\infty$, which
implies that there exists no ground state of the minimization
problem (\ref{groundstate}). $\Box$

By splitting the total energy $E(\cdot)$ in (\ref{newener}) into
kinetic, potential, interaction and dipolar energies, i.e. \be
E(\phi)=E_{\rm kin}(\phi)+E_{\rm pot}(\phi)+E_{\rm int}(\phi)+E_{\rm
dip}(\phi),\ee where \bea  E_{\rm
kin}(\phi)&=&\frac{1}{2}\int_{{\Bbb R}^3} |\nabla\phi(\bx) |^2 d
\bx,\   E_{\rm pot}(\phi) = \int_{{\Bbb R}^3} V(\bx)|\phi(\bx)|^2 d
\bx, \ E_{\rm int}(\phi) =
\frac{\beta}{2}\int_{{\Bbb R}^3} |\phi(\bx) |^4 d \bx,\nn \\
E_{\rm dip} (\phi)&=&\frac{\lambda}{2}\int_{{\Bbb R}^3} \left(U_{\rm
dip}\ast|\phi|^2\right) |\phi(\bx)|^2 d \bx=
\frac{\lambda}{2}\int_{{\Bbb R}^3} |\phi(\bx)|^2\left[3
 \left(\p_{\bf n\bn} \varphi\right)^2-|\phi(\bx)|^2\right]d \bx\nn\\
&=&\frac{\lambda}{2}\int_{{\Bbb R}^3} \left[-|\phi(\bx)|^4+3
\left|\p_{\bf n}\nabla \varphi\right|^2\right]d \bx, \label{dipp03}
 \eea
with $\varphi$ defined in (\ref{poi101}), we have the following
Viral identity:

\begin{proposition}
Suppose $\phi_e$ is a stationary state of a dipolar BEC, i.e. an
eigenfunction of the nonlinear eigenvalue problem (\ref{gpe22dstat})
under the constraint (\ref{const}), then we have \be \label{virial}
2E_{\rm kin}(\phi_e)- 2E_{\rm trap}(\phi_e)+3E_{\rm
int}(\phi_e)+3E_{\rm dip}(\phi_e)=0.\ee
\end{proposition}

\noindent {\bf{Proof:}} Follow the analogous proof for a BEC without
dipolar interaction  \cite{Pitaevskii} and we omit the details here
for brevity. $\Box$

\subsection{Analytical results for dynamics}

The well-posedness of the Cauchy problem of (\ref{ngpe}) was
discussed in \cite{Carles} by analyzing the convolution kernel
$U_{\rm dip}(\bx)$ with detailed Fourier transform. Under the new
formulation (\ref{gpe})-(\ref{poisson}), here we present a simpler
proof for the well-posedness and show finite time blow-up for the
Cauchy problem of a dipolar BEC in different parameter regimes.
Denote
$$ X=\left\{u\in
H^1({\Bbb R}^3)\ \big|\ \|u\|_X^2=\|u\|_{L^2}^2+\|\nabla
u\|_{L^2}^2+\int_{\Bbb R^3}V(\bx)|u(\bx)|^2\,d\bx<\infty\right\}.$$

\begin{theorem} (Well-posedness) Suppose the real-valued trap
potential $V(\bx)\in C^\infty(\Bbb R^3)$ such that $V(\bx)\ge0$ for
$\bx\in{\Bbb R}^3$ and $D^\alpha V(\bx)\in L^\infty(\Bbb R^3)$ for
all $\alpha\in{\Bbb N}_0^3$ with $|\alpha|\ge 2$. For any initial
data $\psi(\bx,t=0)=\psi_0(\bx)\in X$,
 there exists
$T_{\mbox{\rm max}}\in(0,+\infty]$ such that the problem
(\ref{gpe})-(\ref{poisson})
 has a unique maximal solution
$\psi\in C\left([0,T_{\mbox{max}}),X\right)$. It is maximal in the
sense that if $T_{\mbox{\rm max}}<\infty$, then
$\|\psi(\cdot,t)\|_X\to\infty$ when  $t\to T^-_{\mbox{\rm max}}$.
Moreover, the {\sl mass} $N(\psi(\cdot,t))$ and {\sl energy}
$E(\psi(\cdot,t))$ defined in (\ref{norm}) and (\ref{energy}),
respectively, are conserved for $t\in[0,T_{\rm max})$. Specifically,
if $\beta\ge0$ and $-\frac12\beta\leq\lambda\leq\beta$, the solution
to (\ref{gpe})-(\ref{poisson}) is global in time, i.e.,
  $T_{\mbox{max}}=\infty$.
\end{theorem}

\noindent {\bf{Proof: }} For any $\phi\in X$, let $\varphi$ be the
solution of the Poisson equation (\ref{poi11}), denote
$\rho=|\phi|^2$ and define \be
G(\phi,\bar{\phi}):=G(\rho)=\frac12\int_{\Bbb
R^3}|\phi(\bx)|^2\p_{\bf nn}\varphi(\bx)\,d\bx, \qquad
g(\phi)=\frac{\delta G(\phi,\bar{\phi})}{\delta
\bar{\phi}}=\phi\;\p_{\bf nn}\varphi,\qquad \ee where $\bar{f}$
denotes the conjugate of $f$.
 Noticing (\ref{pbnf1}), it is easy to show that
 $G(\phi)\in C^1(X,\Bbb
R)$, $g(\phi) \in C(X,L^p)$ for some $p\in(6/5,2]$, and
\be\label{con3}\|g(u)-g(v)\|_{L^p}\leq
C(\|u\|_X+\|v\|_X)\|u-v\|_{L^r},\quad \hbox{for some }r\in[2,6),
\qquad \forall u,v\in X.\ee Applying the standard Theorems 9.2.1,
4.12.1 and 5.7.1 in \cite{Cazen,Sulem} for the well-posedness of the
nonlinear Schr\"{o}dinger equation, we can obtain the results
immediately. $\Box$

\begin{theorem}(Finite time blow-up) If $\beta<0$, or $\beta\ge0$ and
$\lambda<-\frac{1}{2}\beta$ or $\lambda>\beta$,
 and assume $V(\bx)$ satisfies $3V(\bx)+ \bx\cdot \nabla V(\bx)\ge0$ for
$\bx\in{\Bbb R}^3$. For any initial data
$\psi(\bx,t=0)=\psi_0(\bx)\in X$ to the problem
(\ref{gpe})-(\ref{poisson}), there exists finite time blow-up, i.e.,
$T_{\mbox{max}}<\infty$, if one of the following holds:

(i) $E(\psi_0)<0$;

(ii) $E(\psi_0)=0$ and ${\rm Im}\left(\int_{\Bbb
R^3}\bar{\psi}_0(\bx)\ (\bx\cdot\nabla\psi_0(\bx))\,d\bx\right)<0$;

(iii) $E(\psi_0)>0$ and ${\rm Im}\left(\int_{\Bbb R^3}
\bar{\psi}_0(\bx)\ (\bx\cdot\nabla\psi_0(\bx))\,d\bx\right)
<-\sqrt{3E(\psi_0)}\|\bx\psi_0\|_{L^2}$;

\noindent where  {\rm Im}(f) denotes the  imaginary part of $f$.
\end{theorem}

\noindent {\bf{ Proof :}} Define the variance \be\label{dtv001}
\sg_V(t):=\sg_V(\psi(\cdot,t))=\int_{\Bbb
R^3}|\bx|^2|\psi(\bx,t)|^2\,d\bx=\delta_x(t)+\delta_y(t)+\delta_z(t),\qquad
t\ge0,\ee where \be
\label{dtap01}\sg_\alpha(t):=\sg_\alpha(\psi(\cdot,t))=\int_{\Bbb
R^3}\alpha^2|\psi(\bx,t)|^2\,d\bx, \qquad \alpha=x,\ y,\ z.\ee For
$\alpha=x$, or $y$ or $z$, differentiating (\ref{dtap01}) with
respect to $t$, noticing (\ref{gpe}) and (\ref{poisson}),
integrating by parts, we get \be
\frac{d}{dt}\sg_\alpha(t)=-i\int_{\Bbb
R^3}\left[\alpha\bar{\psi}(\bx,t)\p_{\alpha}\psi(\bx,t)-
\alpha\psi(\bx,t)\p_{\alpha}\bar{\psi}(\bx,t)\right]\,d\bx, \qquad
t\ge0.\ee Similarly, we have \be \label{d2ap22}
\frac{d^2}{dt^2}\sg_\alpha(t)=\int_{\Bbb
R^3}\left[2|\p_{\alpha}\psi|^2+(\beta-\lambda)
|\psi|^4+6\lambda|\psi|^2\alpha\p_{\alpha}\p_{\bf{nn}}
\varphi-2\alpha|\psi|^2\p_{\alpha}V(\bx)\right]\,d\bx. \ee Noticing
(\ref{poisson}) and
\[
-\int_{\Bbb R^3}\nabla^2\varphi\left( \bx\cdot \nabla
\p_{\bf{nn}}\varphi\right)\,d\bx=\frac 32\int_{\Bbb
R^3}|\p_{\bf{n}}\nabla \varphi|^2\,d\bx,
\]
summing (\ref{d2ap22}) for $\alpha=x$, $y$ and $z$, using
(\ref{dtv001}) and (\ref{energy}), we get
\begin{eqnarray}
\frac{d^2}{dt^2}\sg_V(t)&=&2\int_{\Bbb
R^3}\left(|\nabla\psi|^2+\frac 32(\beta-\lambda)|\psi|^4+\frac
92\lambda|\p_{\bf{n}}\nabla\psi|^2-|\psi|^2(\bx\cdot\nabla
V(\bx))\right)\,d\bx\nn\\
&=&6E(\psi)-\int_{\Bbb R^3}|\nabla\psi(\bx,t)|^2-2\int_{\Bbb
R^3}|\psi(\bx,t)|^2\left(3V(\bx)+\bx\cdot\nabla V(\bx)\right)\,d\bx\nn\\
&\leq&6E(\psi)\equiv 6E(\psi_0), \qquad t\ge0.
\end{eqnarray}
Thus,
$$
\sg_V(t)\leq 3E(\psi_0)t^2+\sg_V^\prime(0)t+\sg_V(0), \qquad t\ge0,
$$
and the conclusion follows in the same manner as those in
\cite{Sulem,Cazen} for the standard nonlinear Schr\"{o}dinger
equation. $\Box$

\section{A numerical method for computing ground states} \label{0tssp}
 \setcounter{equation}{0}

Based on the new mathematical formulation for the energy in
(\ref{newener}), we will present an efficient and accurate backward
Euler sine pseudospectral method for computing the ground states of
a dipolar BEC.

In practice, the whole space problem is usually truncated into a
bounded computational domain $\Og=[a,b]\tm[c,d]\tm[e,f]$ with
homogeneous  Dirichlet boundary condition. Various numerical methods
have been proposed in the literatures for computing the ground
states of BEC (see \cite{Schnerder,Tosi,Bao1,Bao3,Bao5,Chang1,Ca1}
and references therein). One of the popular and efficient techniques
for dealing with the constraint (\ref{const}) is through the
following construction \cite{Bao1,Bao2,Bao3}: Choose a time step
$\Delta t>0$ and set $t_n=n\; \Delta t$ for $n=0,1,\ldots$  Applying
the steepest decent method to the energy functional $E(\phi)$ in
(\ref{newener}) without the constraint (\ref{const}), and then
projecting the solution back to the unit sphere $S$ at the end of
each time interval $[t_n,t_{n+1}]$ in order to satisfy the
constraint (\ref{const}). This procedure leads to the function
$\phi(\bx,t)$ is the solution of the following gradient flow with
discrete normalization: \bea \label{ngf1} &&\p_t
\phi(\bx,t)=\left[\fl{1}{2 }\btd^2
-V(\bx)-(\beta-\lambda)|\phi(\bx,t)|^2+
3\lambda \p_{\bf{nn}}\varphi(\bx,t)\right]\phi(\bx,t),    \\
\label{ngf21}&&\nabla^2 \varphi(\bx,t) = -|\phi(\bx,t)|^2, \qquad
\qquad \bx\in\Og,\quad  t_n \leq t < t_{n+1}, \\
\label{ngf2} &&\phi(\bx,t_{n+1}):=
\phi(\bx,t_{n+1}^+)=\fl{\phi(\bx,t_{n+1}^-)}{\|\phi(\cdot,t_{n+1}^-)\|},
\qquad \bx\in \Og, \quad n\ge 0,\\
&&\left.\phi(\bx,t)\right|_{\bx\in\p\Og}=\left.\varphi(\bx,t)\right|_{\bx\in\p\Og}=0,
\qquad t\ge0,\\
 \label{ngf3}
 &&\phi(\bx,0)=\phi_0(\bx), \qquad
\qquad \hbox{with}\quad \|\phi_0\|=1; \eea where
 $\phi(\bx,
t_n^\pm)=\lim\limits_{t\to t_n^\pm} \phi(\bx,t)$.

Let $M$, $K$ and $L$ be even positive integers and define the index
sets \beas &&{\cal T}_{MKL}=\{(j,k,l)\ |\ j=1,2,\ldots,M-1,\
k=1,2,\ldots, K-1, \ l=1,2,\ldots,L-1\}, \\
&&{\cal T}_{MKL}^0=\{(j,k,l)\ |\ j=0,1,\ldots,M,\ k=0,1,\ldots, K, \
l=0,1,\ldots,L\}. \eeas Choose the spatial mesh sizes as
$h_x=\frac{b-a}{M}$, $h_y=\frac{d-c}{K}$ and $h_z=\frac{f-e}{L}$ and
define
\[x_j:=a+j\;h_x,\qquad  y_k = c+ k\; h_y,\qquad
z_l = e+ l\; h_z, \qquad (j,k,l)\in {\cal T}^0_{MKL}.\] Denote the
space
\[Y_{MKL}={\rm
span}\{\Phi_{jkl}(\bx), \quad (j,k,l)\in{\cal T}_{MKL}\},\] with
\[
\Phi_{jkl}(\bx)=\sin\left(\mu_j^x(x-a)\right)\sin\left(\mu_k^y(y-c)\right)
\sin\left(\mu_l^z(z-e)\right), \quad \bx\in \Og,\qquad
(j,k,l)\in{\cal T}_{MKL}, \]
\[\mu_j^x =
\fl{\pi j}{b-a}, \qquad \mu_k^y = \fl{\pi k}{d-c}, \qquad \mu_l^z =
\fl{\pi l}{f-e}, \qquad (j,k,l)\in{\cal T}_{MKL}; \] and  $P_{MKL}:
Y=\{\varphi\in C(\Og)\ |\ \varphi(\bx)|_{\bx\in\p\Og}=0\}\to
Y_{MKL}$ be the standard project operator \cite{ST}, i.e.
\[(P_{MKL}v)(\bx)=\sum_{p=1}^{M-1}\sum_{q=1}^{K-1}\sum_{s=1}^{L-1}
\widehat{v}_{pqs}\; \Phi_{pqs}(\bx), \quad \bx\in\Og,\qquad \forall
v\in Y,
\]
with \be\label{FST} \widehat{v}_{pqs}=\int_{\Og} v(\bx)\;
\Phi_{pqs}(\bx)\;d\bx, \qquad (p,q,s)\in{\cal T}_{MKL}. \ee Then a
backward Euler sine spectral discretization
for (\ref{ngf1})-(\ref{ngf3}) reads:\\
 Find
$\phi^{n+1}(\bx)\in Y_{MKL}$ (i.e. $\phi^{+}(\bx)\in Y_{MKL}$) and
$\varphi^{n}(\bx)\in Y_{MKL}$ such that {\small \bea
&&\frac{\phi^{+}(\bx)-\phi^n(\bx)}{\Delta t}=\frac{1}{2}\nabla^2
\phi^{+}(\bx)
-P_{MKL}\left\{\left[V(\bx)+(\beta-\lambda)|\phi^n(\bx)|^2+ 3\lambda
\p_{\bf{nn}}\varphi^n(\bx)\right]\phi^{+}(\bx)\right\},\qquad  \\
&&\nabla^2\varphi^n(\bx)=-P_{MKL}\left(|\phi^n(\bx)|^2\right),\qquad
\phi^{n+1}(\bx)=\frac{\phi^{+}(\bx)}{\|\phi^{+}(\bx)\|_2}, \qquad
\bx\in\Og,\quad  n\ge0; \eea} where
$\phi^0(\bx)=P_{MKL}\left(\phi_0(\bx)\right)$ is given.

The above discretization can be solved in phase space and it is not
suitable in practice due to the difficulty of computing the
integrals in (\ref{FST}). We now present an efficient implementation
by choosing $\phi^0(\bx)$ as the interpolation of $\phi_0(\bx)$ on
the grid points $\{(x_j,y_k,z_l), \ (j,k,l)\in{\cal T}_{MKL}^0\}$,
i.e $\phi^0(x_j,y_k,z_l) =\phi_0(x_j,y_k,z_l)$ for $(j,k,l)\in{\cal
T}_{MKL}^0$, and approximating the integrals in (\ref{FST}) by a
quadrature rule on the grid points. Let $\phi_{jkl}^n$ and
$\varphi_{jkl}^n$ be the approximations of $\phi(x_j,y_k,z_l,t_n)$
and $\varphi(x_j,y_k,z_l,t_n)$, respectively, which are the solution
of (\ref{ngf1})-(\ref{ngf3}); denote $\rho_{jkl}^n=|\phi^n_{jkl}|^2$
and  choose $\phi_{jkl}^0=\phi_0(x_j,y_k,z_l)$ for $(j,k,l)\in {\cal
T}_{MKL}^0$. For $n=0,1,\ldots$, a backward Euler sine
pseduospectral discretization for (\ref{ngf1})-(\ref{ngf3}) reads:
{\small \bea &&\fl{\phi_{jkl}^+-\phi_{jkl}^n}{\triangle t}=
\fl{1}{2} \left.\left(\nabla_s^2
\phi^+\right)\right|_{jkl}-\left[V(x_j,y_k,z_l)+(\beta-\lambda)
\left|\phi_{jkl}^n\right|^2 +3\lambda\left.\left(\p_{\bn\bn}^s
\varphi^n\right)\right|_{jkl}\right] \phi^+_{jkl}, \qquad
\label{discretized1} \\
&&-\left.\left(\nabla_s^2 \varphi^n\right)\right|_{jkl}=
|\phi_{j,k,l}^n|^2=\rho_{jkl}^n, \qquad
\phi_{jkl}^{n+1}=\fl{\phi_{jkl}^+}{\|\phi^+\|_h}, \qquad (j,k,l)\in
{\cal T}_{MKL}, \\
&&\phi_{0kl}^{n+1}=\phi_{Mkl}^{n+1}=\phi_{j0l}^{n+1}=
\phi_{jKl}^{n+1}=\phi_{jk0}^{n+1}=\phi_{jkL}^{n+1}=0,\qquad
(j,k,l)\in {\cal T}_{MKL}^0,\\
&&\varphi_{0kl}^{n}
=\varphi_{Mkl}^{n}=\varphi_{j0l}^{n}=\varphi_{jKl}^{n}=
\varphi_{jk0}^{n}=\varphi_{jkL}^{n}=0, \qquad (j,k,l)\in {\cal
T}_{MKL}^0;\label{discretized2} \eea}  where $\nabla_s^2$ and
$\p_{\bn\bn}^s$ are sine pseudospectral
 approximations of $\nabla^2$ and $\p_{\bn\bn}$, respectively,
 defined as
{\small \bea&&\left.\left(\nabla_s^2 \phi^n\right)\right|_{jkl} =
-\sum_{p=1}^{M-1}\sum_{q=1}^{K-1}\sum_{s=1}^{L-1}
\left[(\mu_p^x)^2+(\mu_q^y)^2+(\mu_s^z)^2\right]\widetilde{(\phi^n)}_{pqs}
\sin\left(\frac{jp\pi}{M}\right)\sin\left(\frac{kq\pi}{K}\right)
\sin\left(\frac{ls\pi}{L}\right),\qquad \nn \\
&&\left.\left(\p_{\bn\bn}^s
\varphi^n\right)\right|_{jkl}=\sum_{p=1}^{M-1}\sum_{q=1}^{K-1}\sum_{s=1}^{L-1}
\frac{\widetilde{(\rho^n)}_{pqs}}{(\mu_p^x)^2+(\mu_q^y)^2+(\mu_s^z)^2}
\left.\left(\p_{\bn\bn}\Phi_{pqs}(\bx)\right)\right|_{(x_j,y_k,z_l)},
\ (j,k,l)\in {\cal T}_{MKL}, \label{dstpp}
 \eea}
with $\widetilde{(\phi^n)}_{pqs}$ ($(p,q,s)\in{\cal T}_{MKL})$ the
discrete sine transform coefficients of the vector $\phi^n$ as
{\small \be\label{dst11}\widetilde{(\phi^n)}_{pqs}=
\frac{8}{MKL}\sum_{j=1}^{M-1}\sum_{k=1}^{K-1}\sum_{l=1}^{L-1}
\phi^n_{jkl}
\sin\left(\frac{jp\pi}{M}\right)\sin\left(\frac{kq\pi}{K}\right)
\sin\left(\frac{ls\pi}{L}\right), \quad (p,q,s)\in {\cal
T}_{MKl},\ee } and the discrete $h$-norm is defined as
\[ \|\phi^+\|_h^2 = h_xh_yh_z\sum_{j=1}^{M-1}\sum_{k=1}^{N-1}\sum_{l=1}^{L-1}
|\phi_{jkl}^+|^2.\] Similar as those in \cite{Bao6}, the linear
system (\ref{discretized1})-(\ref{discretized2}) can be iteratively
solved in phase space very efficiently via discrete sine transform
and we omitted the details here for brevity.

\section{A time-splitting sine pseudospectral method for dynamics}\label{1tssp}

Similarly, based on the new Gross-Pitaevskii-Poisson type system
(\ref{gpe})-(\ref{poisson}), we will present an efficient and
accurate time-splitting sine pseudospectral (TSSP) method for
computing the dynamics of a dipolar BEC.

Again, in practice, the whole space problem is truncated into a
bounded computational domain $\Og=[a,b]\tm[c,d]\tm[e,f]$ with
homogeneous  Dirichlet boundary condition. From time $t=t_n$ to time
$t=t_{n+1}$, the Gross-Pitaevskii-Poisson type system
(\ref{gpe})-(\ref{poisson}) is solved in two steps.  One solves
first \be\label{fgpe}i \p_t
\psi(\bx,t)=-\fl{1}{2}\nabla^2\psi(\bx,t), \quad \bx\in\Og, \qquad
\left.\psi(\bx,t)\right|_{\bx\in\p\Og}=0, \qquad t_n\le t\le
t_{n+1},\ee for the time step of length $\Delta t$, followed by
solving \bea\label{ode11} &&i \p_t
\psi(\bx,t)=\left[V(\bx)+(\beta-\lambda) |\psi(\bx,t)|^2-3\lambda
\p_{\bn\bn} \varphi(\bx,t)
\right]\psi(\bx,t),  \\
\label{poisson11}&&\nabla^2 \varphi(\bx,t) = -|\psi(\bx,t)|^2,\qquad
 \bx\in\Og, \qquad t_n\le t \le t_{n+1}; \\ \label{bond123}
 &&\left.\varphi(\bx,t)\right|_{\bx\in\p\Og}=0, \qquad \left.\psi(\bx,t)\right|_{\bx\in\p\Og}=0,
 \qquad t_n\le t \le t_{n+1};\eea
for the same time step. Equation (\ref{fgpe}) will be discretized in
space by sine pseudospectral method and integrated in time {\sl
exactly} \cite{Bao8}. For $t\in[t_n,t_{n+1}]$, the equations
(\ref{ode11})-(\ref{bond123}) leave $|\psi|$ and $\varphi$ invariant
in $t$ \cite{Bao_Jaksch_Markowich,Bao8} and therefore they collapses
to {\small \bea\label{ode111} &&i \p_t
\psi(\bx,t)=\left[V(\bx)+(\beta-\lambda) |\psi(\bx,t_n)|^2-3\lambda
\p_{\bn\bn} \varphi(\bx,t_n)
\right]\psi(\bx,t),  \quad \bx\in\Og,\ t_n\le t \le t_{n+1},\qquad \quad  \\
\label{poisson113}&&\nabla^2 \varphi(\bx,t_n) =
-|\psi(\bx,t_n)|^2,\qquad
 \bx\in\Og.
\eea} Again, equation (\ref{poisson113}) will be discretized in
space by sine pseudospectral method \cite{Bao8,ST} and the linear
ODE (\ref{ode111}) can be integrated in time {\sl exactly}
\cite{Bao_Jaksch_Markowich,Bao8}.

Let $\psi_{jkl}^n$ and $\varphi_{jkl}^n$ be the approximations of
$\psi(x_j,y_k,z_l,t_n)$ and $\varphi(x_j,y_k,z_l,t_n)$,
respectively, which are the solution of (\ref{gpe})-(\ref{poisson});
and choose $\psi^0_{jkl}=\psi_0(x_j,y_k,z_l)$ for $(j,k,l)\in {\cal
T}_{MKL}^0$. For $n=0,1,\ldots$,  a second-order TSSP method for
solving (\ref{gpe})-(\ref{poisson}) via the standard Strang
splitting is \cite{str,Bao_Jaksch_Markowich,Bao8} {\small
\bea\label{tssp1}
 &&\psi^{(1)}_{jkl}=\sum_{p=1}^{M-1}\sum_{q=1}^{K-1}\sum_{s=1}^{L-1}
e^{-i\triangle t\left[(\mu_p^x)^2+(\mu_q^y)^2+(\mu_r^z)^2 \right]
/4}\;\widetilde{(\psi^n)}_{pqr}\sin\left(\frac{jp\pi}{M}
\right)\sin\left(\frac{kq\pi}{K}\right)
\sin\left(\frac{ls\pi}{L}\right), \nn \\
 &&\psi^{(2)}_{jkl}=
 e^{-i\triangle t\left[V(x_j,y_k,z_l)+(\bt-\lambda)|\psi_{jkl}^{(1)}|^2-3\lambda
\left.\left(\p_{\bn\bn}^s\varphi^{(1)}\right)\right|_{jkl}\right]
}\; \psi_{jkl}^{(1)},
\qquad (j,k,l)\in{\cal T}_{MKL}^0, \\
&&\psi^{n+1}_{jkl}=\sum_{p=1}^{M-1}\sum_{q=1}^{K-1}\sum_{s=1}^{L-1}
e^{-i\triangle t\left[(\mu_p^x)^2+(\mu_q^y)^2+(\mu_r^z)^2 \right]
/4}\;\widetilde{(\psi^{(2)})}_{pqr}\sin\left(\frac{jp\pi}{M}
\right)\sin\left(\frac{kq\pi}{K}\right)
\sin\left(\frac{ls\pi}{L}\right); \nn
 \eea}
where $\widetilde{(\psi^n)}_{pqr}$ and
$\widetilde{(\psi^{(2)})}_{pqr}$ ($(p,q,s)\in{\cal T}_{MKL}$) are
the discrete sine transform coefficients of the vectors $\psi^n$ and
$\psi^{(2)}$, respectively (defined similar as those in
(\ref{dst11})); and
$\left.\left(\p_{\bn\bn}^s\varphi^{(1)}\right)\right|_{jkl}$
 can be computed as in (\ref{dstpp}) with
$\rho^n_{jkl}=\rho^{(1)}_{jkl}:=|\psi^{(1)}_{jkl}|^2$  for
$(j,k,l)\in {\cal T}_{MKL}^0$.

The above method is explicit, unconditionally stable, the memory
cost is $O(MKL)$ and the computational cost per time step is
$O\left(MKL\ln(MKL)\right)$. In fact, for the stability, we have

\begin{lemma}
The TSSP method (\ref{tssp1}) is normalization conservation, i.e.
\bea
\|\psi^n\|_h^2:=h_xh_yh_z\sum_{j=1}^{M-1}\sum_{k=1}^{K-1}\sum_{l=1}^{L-1}|\psi^n_{jkl}|^2\equiv
h_xh_yh_z\sum_{j=1}^{M-1}\sum_{k=1}^{K-1}\sum_{l=1}^{L-1}|\psi^0_{jkl}|^2=\|\psi^0\|_h^2,
\ n\ge0. \quad \eea

\end{lemma}

\noindent {\bf Proof:} Follow the analogous proof in
\cite{Bao_Jaksch_Markowich,Bao8} and we omit the details here for
brevity. $\Box$

\section{Numerical results}
\setcounter{equation}{0} In this section, we first compare our new
methods and the standard method used in the literatures
\cite{Yi2,Xiong,Tick,Blakie} to evaluate numerically the dipolar
energy and then report ground states and dynamics of dipolar BECs by
using our new numerical methods.

\subsection{Comparison for evaluating the dipolar energy}

Let \be \phi:=\phi(\bx) = \pi^{-3/4} \gamma_{x}^{1/2} \gamma_z^{1/4}
e^{-\fl{1}{2}\left( \gamma_x (x^2+y^2)+\gamma_z z^2\right)},\qquad
\bx\in{\Bbb R}^3.\ee Then the
 dipolar energy $E_{\rm dip}(\phi)$ in (\ref{dipp03})
 can be  evaluated analytically as \cite{Tikhonenkov}
\be E_{\rm dip}(\phi)= -\fl{\lambda\gamma_x \sqrt{\gamma_z}}{4\pi
\sqrt{2\pi} }\left\{
\begin{array}{ll}
  \fl{1+2\kappa^2}{1-\kappa^2}-\fl{3\kappa^2 \rm{arctan} \left(
\sqrt{\kappa^2-1}\right)}{(1-\kappa^2)\sqrt{\kappa^2-1}}, & \kappa>1, \\
  0, & \kappa =1, \\
  \fl{1+2\kappa^2}{1-\kappa^2}-\fl{1.5\kappa^2
}{(1-\kappa^2)\sqrt{1-\kappa^2}} \rm{ln} \left(
\fl{1+\sqrt{1-\kappa^2}}{1-\sqrt{1-\kappa^2}}\right), & \kappa <1, \\
\end{array}
\right. \ee
 with $\kappa =\sqrt{ \fl{\gamma_z}{\gamma_x}}$. This provides a
 perfect example to test the efficiency of different numerical
 methods to deal with the dipolar potential. Based on our new
 formulation (\ref{dipp03}), the dipolar energy can be evaluated via
 discrete sine transform (DST) as
\bea \label{dstdip1}E_{\rm dip}(\phi)\approx \frac{\lambda
h_xh_yh_z}{2}
\sum_{j=1}^{M-1}\sum_{k=1}^{K-1}\sum_{l=1}^{L-1}|\phi(x_j,y_k,z_l)|^2\left[
3\left(\left.\left(\p_{\bn\bn}^s\varphi^n
\right)\right|_{jkl}\right)^2-|\phi(x_j,y_k,z_l)|^2\right], \nn\eea
where $\left.\left(\p_{\bn\bn}^s\varphi^n \right)\right|_{jkl}$ is
computed as in (\ref{dstpp}) with
$\rho^n_{jkl}=|\phi(x_j,y_k,z_l)|^2$ for $(j,k,l)\in {\cal \cal
T}_{MKL}^0$. In the literatures \cite{Yi2,Tick,Xiong,Blakie}, this
dipolar energy is usually calculated via discrete Fourier transform
(DFT) as \bea \label{dftdip1}E_{\rm dip}(\phi)\approx \frac{\lambda
h_xh_yh_z}{2}
\sum_{j=1}^{M-1}\sum_{k=1}^{K-1}\sum_{l=1}^{L-1}|\phi(x_j,y_k,z_l)|^2\left[
{\cal F}^{-1}_{jkl}\left(\widehat{(U_{\rm
dip})}(2\mu_p^x,2\mu_q^y,2\mu_s^z)\cdot {\cal
F}_{pqs}(|\phi|^2)\right) \right], \nn\eea where ${\cal F}$ and
${\cal F}^{-1}$ are the discrete Fourier and inverse Fourier
transforms over the grid points $\{(x_j,y_k,z_l), \ (j,k,l)\in {\cal
T}_{MKL}^0\}$, respectively \cite{Xiong}. We take $\lambda=24\pi$,
the bounded computational domain  $\Og=[-16,16]^3$, $M=K=L$ and thus
$h=h_x=h_y=h_z=\frac{32}{M}$. Table \ref{tab1} lists the errors
$e:=\left|E_{\rm dip}(\phi)-E_{\rm dip}^h\right|$ with $E_{\rm
dip}^h$ computed numerically via either (\ref{dstdip1}) or
(\ref{dftdip1}) with mesh size $h$ for three cases:

\begin{itemize}
\item Case I. $\gamma_x=0.25$ and $\gamma_z=1$ which implies $\kappa
=2.0$ and $E_{\rm dip}(\phi) = 0.0386708614$;

\item Case II. $\gamma_x=\gamma_z=1$ which implies
$\kappa =1.0$ and  $E_{\rm dip}(\phi)= 0$;

\item Case III. $\gamma_x=2$ and $\gamma_z=1$ which implies
$\kappa =\sqrt{0.5}$ and  $E_{\rm dip}(\phi)=-0.1386449741$.
\end{itemize}

\begin{table}[htb]
\begin{center}
\begin{tabular}{c|cc|cc|cc}
  \hline
  &\multicolumn{2}{c|}{Case I}&\multicolumn{2}{c|}{Case II}&\multicolumn{2}{c}{Case III}\\ \cline{2-7}
  &DST &DFT &DST &DFT &DST &DFT\\\hline
 $M=32 \& h=1$ &2.756E-2 &2.756E-2 &3.555E-18 &1.279E-4 &0.1018
 &0.1020\\
 $M=64 \& h=0.5$ &1.629E-3 &1.614E-3  &9.154E-18 &1.278E-4
 &9.788E-5
 &2.269E-4\\
 $M=128 \& h=0.25$ &1.243E-7  &1.588E-5 &7.454E-17 &1.278E-4
&6.406E-7 &1.284E-4  \\
  \hline
\end{tabular}
\end{center}
\caption{Comparison for evaluating dipolar energy under different
mesh sizes $h$.} \label{tab1}
\end{table}

From Tab. \ref{tab1} and our extensive numerical results not shown
here for brevity, we can conclude that our new method via discrete
sine transform based on a new formulation  is much more accurate
than that of the standard method via discrete Fourier transform in
the literatures for evaluating the dipolar energy.

\subsection{Ground states of dipolar BECs}

By using our new numerical method
(\ref{discretized1})-(\ref{discretized2}), here we report the ground
states of a dipolar BEC (e.g., ${}^{52}$Cr \cite{Parker}) with
different parameters and trapping potentials. In our computation and
results, we always use the dimensionless quantities. We take
$M=K=L=128$, time step $\Delta t=0.01$, dipolar direction
$\bn=(0,0,1)^T$ and the bounded computational domain $\Og=[-8,8]^3$
for all cases except $\Og=[-16,16]^3$ for the cases
$\frac{N}{10000}=1,\;5,\;10$ and $\Og=[-20,20]^3$ for the cases
$\frac{N}{10000}=50,\;100$ in Table \ref{tab2}. The ground state
$\phi_g$ is reached numerically when
$\|\phi^{n+1}-\phi^n\|_\infty:=\max\limits_{0\le j\le M,\ 0\le k\le
K,\ 0\le l\le L} |\phi^{n+1}_{jkl}-\phi^n_{jkl}|\le \vep:=10^{-6}$
in (\ref{discretized1})-(\ref{discretized2}). Table \ref{tab2} shows
the energy $E^g:=E(\phi_g)$, chemical potential
$\mu^g:=\mu(\phi_g)$, kinetic energy $E_{\rm kin}^g:=E_{\rm
kin}(\phi_g)$, potential energy $E_{\rm pot}^g:=E_{\rm
pot}(\phi_g)$, interaction energy $E_{\rm int}^g:=E_{\rm
int}(\phi_g)$, dipolar energy $E_{\rm dip}^g:=E_{\rm dip}(\phi_g)$,
condensate widths $\sg_x^g:=\sigma_x(\phi_g)$ and
$\sg_z^g:=\sigma_z(\phi_g)$ in (\ref{dtap01}) and central density
$\rho_g({\bf 0}):=|\phi_g(0,0,0)|^2$ with harmonic potential
$V(x,y,z)= \fl{1}{2}\left(x^2+y^2+0.25z^2\right)$ for different
$\beta=0.20716N$ and $\lambda=0.033146N$ with $N$ the total number
of particles in the condensate; and Table \ref{tab3} lists similar
results  with $\beta=207.16$ for different values of $-0.5\le
\frac{\lambda}{\beta}\le 1$. In addition, Figure \ref{fig1} depicts
the ground state $\phi_g(\bx)$, e.g. surface plots of
$|\phi_g(x,0,z)|^2$ and isosurface plots of  $|\phi_g(\bx)|=0.01$,
 of a dipolar BEC with $\beta = 401.432$ and $\lambda
=0.16\beta$ for harmonic potential $V(\bx)=
\fl{1}{2}\left(x^2+y^2+z^2\right)$, double-well potential
$V(\bx)=\fl{1}{2}\left(x^2+y^2+z^2\right)+4e^{-z^2/2}$ and optical
lattice potential
$V(\bx)=\fl{1}{2}\left(x^2+y^2+z^2\right)+100\left[\sin^2\left(\fl{\pi}{2}x\right)
+\sin^2\left(\fl{\pi}{2}y\right)+\sin^2\left(\fl{\pi}{2}z\right)
\right]$; and Figure \ref{fig6} depicts  the ground state
$\phi_g(\bx)$, e.g. isosurface plots of $|\phi_g(\bx)|=0.08$, of a
dipolar BEC with the harmonic potential $V(\bx)=
\fl{1}{2}\left(x^2+y^2+z^2\right)$ and $\beta=207.16$ for different
values of $-0.5\le \frac{\lambda}{\beta}\le 1$.

\begin{table}[htb]
\begin{center}
\begin{tabular}{cccccccccc}
\hline \\
$\frac{N}{10000}$ &$E^g$ &$\mu^g$  &$E_{\rm kin}^g$ &$E_{\rm pot}^g$
&$E_{\rm int}^g$ &$E_{\rm dip}^g$
&$\sigma_x^g$ &$\sigma_z^g$& $\rho_g({\bf 0})$\\
\hline

0.1 &1.567 &1.813 &0.477 &0.844 &0.262 &-0.015 &0.796 &1.299 &0.06139 \\
0.5 &2.225 &2.837 &0.349 &1.264 &0.659 &-0.047 &0.940 &1.745 &0.02675 \\
1   &2.728 &3.583 &0.296 &1.577 &0.925 &-0.070 &1.035 &2.009 &0.01779\\
5   &4.745 &6.488 &0.195 &2.806 &1.894 &-0.151 &1.354 &2.790 &0.00673 \\
10  &6.147 &8.479 &0.161 &3.654 &2.536 &-0.204 &1.538 &3.212 &0.00442     \\
50  &11.47 &15.98 &0.101 &6.853 &4.909 &-0.398 &2.095 &4.441 &0.00168  \\
100 &15.07 &21.04 &0.082 &9.017 &6.498  &-0.526 &2.400 &5.103 &0.00111  \\

\hline
\end{tabular}
 \end{center}
\caption{Different quantities of the ground states of a dipolar BEC
 for $\beta=0.20716N$ and $\lambda=0.033146N$ with different number of particles
$N$.} \label{tab2}
 \end{table}

\begin{table}[htb]
\begin{center}
\begin{tabular}{cccccccccc}
\hline \\
 $\fl{\lambda}{\beta}$  &  $E^g$  & $\mu^g$ & $E_{kin}^g$ &   $E_{pot}^g$ &  $E_{int}^g$
 &  $E_{dip}^g$ &   $\sigma_x^g$ & $\sigma_z^g$ & $\rho_g({\bf 0})$
 \\ \hline

-0.5 &2.957 &3.927 &0.265 &1.721 &0.839 &0.131 &1.153 &1.770 &0.01575 \\
-0.25 &2.883 &3.817 &0.274 &1.675 &0.853 &0.081 &1.111 &1.879 &0.01605 \\
0 &2.794 &3.684 &0.286 &1.618 &0.890 &0.000 &1.066 &1.962 &0.01693 \\
0.25 &2.689 &3.525 &0.303 &1.550 &0.950 &-0.114 &1.017 &2.030 &0.01842 \\
0.5 &2.563 &3.332 &0.327 &1.468 &1.047 &-0.278 &0.960 &2.089 &0.02087 \\
0.75 &2.406 &3.084 &0.364 &1.363 &1.212 &-0.534 &0.889 &2.141 &0.02536\\
1.0 &2.193 &2.726 &0.443 &1.217 &1.575 &-1.041 &0.786  &2.189 &0.03630 \\
 \hline

\end{tabular}
 \end{center}
 \caption{ Different quantities of the ground states of a dipolar BEC with different
 values of $\frac{\lambda}{\beta}$ with $\beta=207.16$.}\label{tab3}
 \end{table}

\begin{figure}[h!]
\centerline{
\psfig{figure=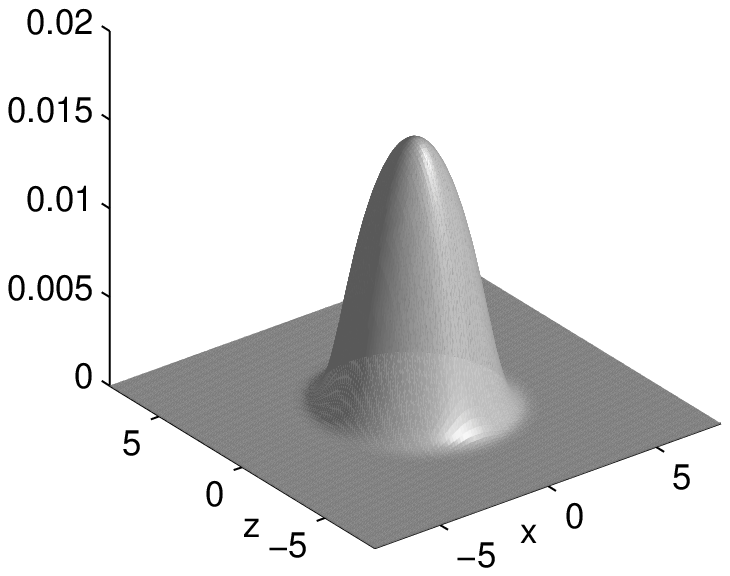,height=6cm,width=6cm,angle=0}\qquad
\psfig{figure=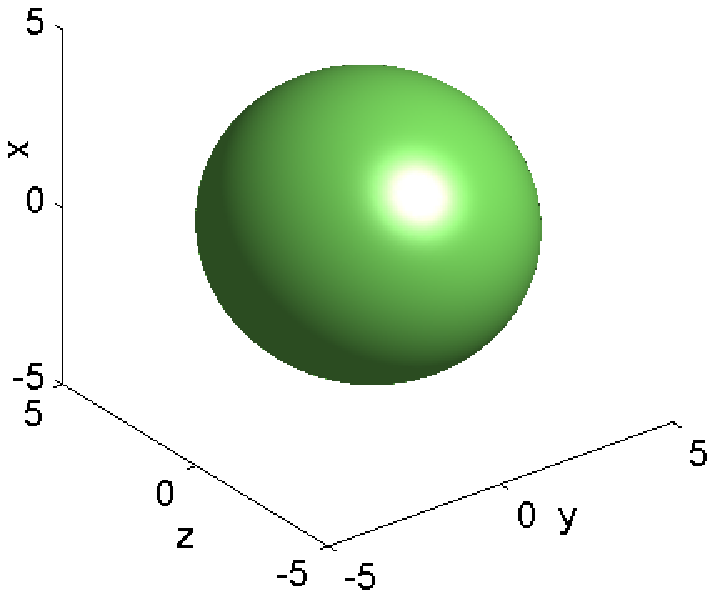,height=6cm,width=6cm,angle=0}
 }
 \centerline{
\psfig{figure=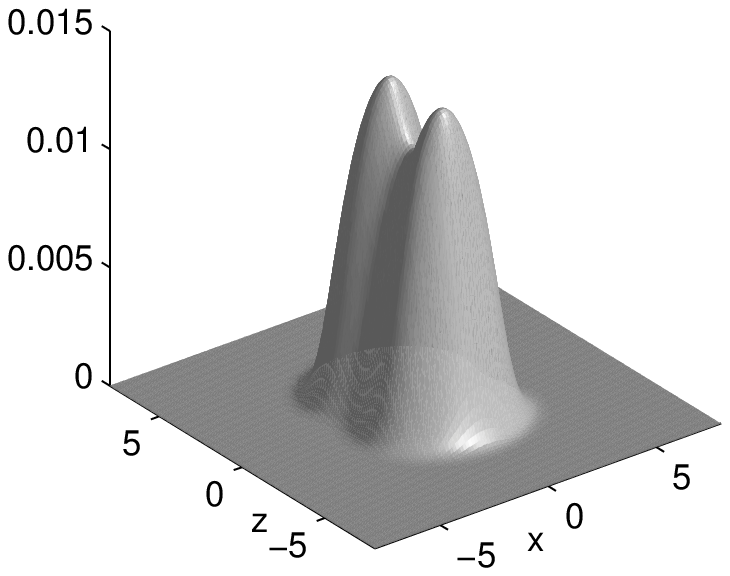,height=6cm,width=6cm,angle=0}\qquad
\psfig{figure=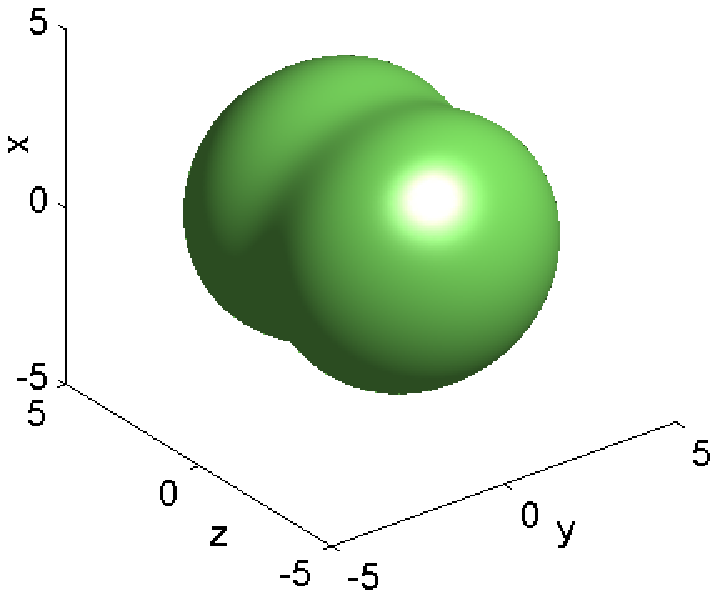,height=6cm,width=6cm,angle=0}
 }
 \centerline{
\psfig{figure=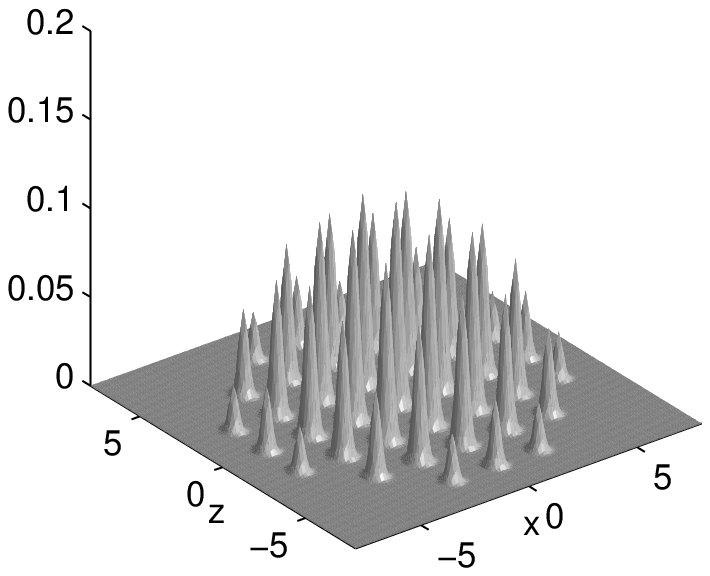,height=6cm,width=6cm,angle=0}\qquad
\psfig{figure=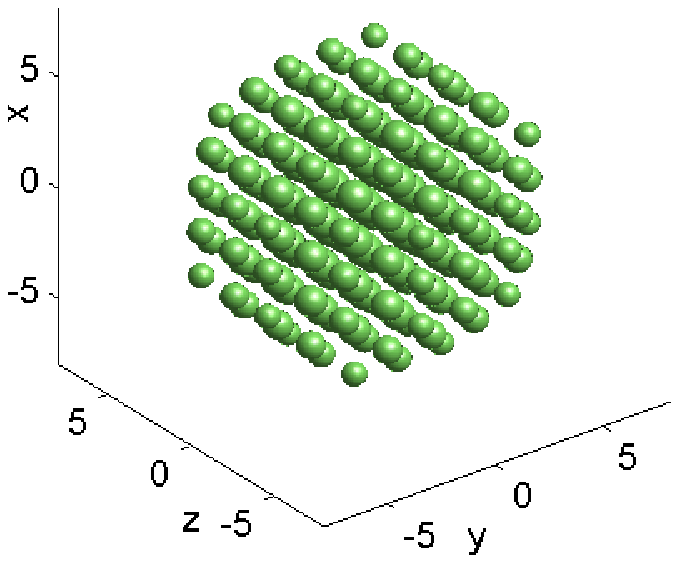,height=6cm,width=6cm,angle=0}
 }
  \caption{
Surface plots of $|\phi_g(x,0,z)|^2$ (left column) and isosurface
plots of $|\phi_g(x,y,z)|=0.01$ (right column) for the ground state
of a dipolar BEC with $\beta= 401.432$ and $\lambda = 0.16 \beta$
for harmonic potential (top row), double-well potential (middle row)
and optical lattice potential (bottom row).
  } \label{fig1}
\end{figure}


From Tabs. \ref{tab2}\&\ref{tab3} and Figs. \ref{fig1}\&\ref{fig6},
we can draw the following conclusions: (i) For fixed trapping
potential $V(\bx)$ and dipolar direction $\bn=(0,0,1)^T$, when
$\beta$ and $\lambda$ increase with the ratio
$\frac{\lambda}{\beta}$ fixed, the energy $E^g$, chemical potential
$\mu^g$,  potential energy $E_{\rm pot}^g$, interaction energy
$E_{\rm int}^g$,  condensate widths $\sg_x^g$ and $\sg_z^g$ of the
ground states increase; and resp., the  kinetic energy $E_{\rm
kin}^g$, dipolar energy $E_{\rm dip}^g$ and central density
$\rho_g({\bf 0})$ decrease (cf. Tab. \ref{tab2}). (ii) For fixed
trapping potential $V(\bx)$, dipolar direction $\bn=(0,0,1)^T$ and
$\beta$, when the ratio $\frac{\lambda}{\beta}$ increases from
$-0.5$ to $1$, the kinetic energy $E_{\rm kin}^g$, interaction
energy $E_{\rm int}^g$, condensate widths $\sg_z^g$ and central
density $\rho_g({\bf 0})$ of the ground states increase; and resp.,
the energy $E^g$, chemical potential $\mu^g$, potential energy
$E_{\rm pot}^g$, dipolar energy $E_{\rm dip}^g$ and condensate
widths $\sg_x^g$
 decrease (cf. Tab. \ref{tab3}). (iii) Our new numerical method can
 compute the ground states accurately and efficiently (cf. Figs.
 \ref{fig1}\&\ref{fig6}).

\subsection{Dynamics of dipolar BECs}

Similarly, by using our new numerical method (\ref{tssp1}), here we
report the dynamics of a dipolar BEC (e.g., ${}^{52}$Cr
\cite{Parker}) under different setups. Again, in our computation and
results, we always use the dimensionless quantities. We take the
bounded computational domain $\Og=[-8,8]^2\times[-4,4]$,
$M=K=L=128$, i.e. $h=h_x=h_y=1/8,h_z=1/16$,  time step $\Delta
t=0.001$. The initial data $\psi(\bx,0)=\psi_0(\bx)$ is chosen as
the ground state of a dipolar BEC computed numerically by our
numerical method with $\bn=(0,0,1)^T$,
$V(\bx)=\fl{1}{2}(x^2+y^2+25z^2)$, $\beta=103.58$ and
$\lambda=0.8\beta=82.864$.

The first case  to study numerically is the dynamics of suddenly
changing the dipolar direction from $\bn=(0,0,1)^T$ to
$\bn=(1,0,0)^T$ at $t=0$ and keeping all other quantities unchanged.
Figure \ref{fig2} depicts time evolution of the energy
$E(t):=E(\psi(\cdot,t))$, chemical potential
$\mu(t)=\mu(\psi(\cdot,t)$, kinetic energy $E_{\rm kin}(t):=E_{\rm
kin}(\psi(\cdot,t))$, potential energy $E_{\rm pot}(t):=E_{\rm
pot}(\psi(\cdot,t))$, interaction energy $E_{\rm int}(t):=E_{\rm
int}(\psi(\cdot,t))$, dipolar energy $E_{\rm dip}(t):=E_{\rm
dip}(\psi(\cdot,t))$, condensate widths
$\sg_x(t):=\sg_x(\psi(\cdot,t))$, 
$\sg_z(t):=\sg_z(\psi(\cdot,t))$, and central density
$\rho(t):=|\psi({\bf 0},t)|^2$, as well as the isosurface of the
density function $\rho(\bx,t):=|\psi(\bx,t)|^2=0.01$ for different
times. In addition, Figure \ref{fig3} show similar results for the
case of
 suddenly
changing the trapping potential   from
$V(\bx)=\fl{1}{2}(x^2+y^2+25z^2)$ to
$V(\bx)=\fl{1}{2}(x^2+y^2+\fl{25}{4}z^2)$ at $t=0$, i.e. decreasing
the trapping frequency in z-direction from $5$ to $\fl{5}{2}$,  and
keeping all other quantities unchanged;  Figure \ref{fig4}  show the
results for the case of
 suddenly
changing the dipolar interaction from $\lambda=0.8\beta=82.864$ to
$\lambda=4\beta=414.32$ at $t=0$ while keeping all other quantities
unchanged, i.e. collapse of a dipolar BEC; and Figure \ref{fig7}
show the results for the case of
 suddenly
changing the interaction constant $\beta$ from $\beta=103.58$ to
$\beta=-569.69$ at $t=0$ while keeping all other quantities
unchanged, i.e. another collapse of a dipolar BEC.

From Figs. \ref{fig2}, \ref{fig3}, \ref{fig4} and \ref{fig7}, we can
conclude that the dynamics of dipolar BEC can be very interesting
and complicated. In fact, global existence of the solution is
observed in the first two cases (cf. Figs. \ref{fig2}\&\ref{fig3})
and finite time blow-up is observed in the last two cases (cf. Figs.
\ref{fig4}\&\ref{fig7}). The total energy is numerically conserved
very well in our computation when there is no blow-up (cf. Figs.
\ref{fig2}\&\ref{fig3}) and before blow-up happens (cf. Figs.
\ref{fig4}\&\ref{fig7}). Of course, it is not conserved numerically
near or after blow-up happens because the mesh size and time step
are fixed which cannot resolve the solution. In addition, our new
numerical method can compute the dynamics of dipolar BEC accurately
and efficiently.

\begin{remark} Due to size limit at ariv, to read the full figures,
you can download this paper from:\\
http://www.math.nus.edu.sg/\~{ }bao/PS/dipolar-bec.pdf
\end{remark}

\section{Conclusions}
\setcounter{equation}{0}

Efficient and accurate numerical methods were proposed for computing
ground states and dynamics of dipolar Bose-Einstein condensates
based on the three-dimensional Gross-Pitaevskii equation (GPE)  with
a nonlocal dipolar interaction potential. By decoupling the dipolar
interaction potential into a short-range and a long-range part, the
GPE for a dipolar BEC is re-formulated to a Gross-Pitaevskii-Poisson
type system. Based on this new mathematical formulation, we proved
rigorously the existence and uniqueness as well as nonexistence of
the ground states, and discussed the dynamical properties of dipolar
BEC in different parameter regimes. In addition, the backward Euler
sine pseudospectral method and time-splitting sine pseudospectral
method were proposed for computing the ground states and dynamics of
a dipolar BEC, respectively. Our new numerical methods avoided
taking the Fourier transform of the nonlocal dipolar interaction
potential which is highly singular and causes some numerical
difficulties in practical computation. Comparison between our new
numerical methods and existing numerical methods in the literatures
showed that our numerical methods perform better. Applications of
our new numerical methods for computing the ground states and
dynamics of dipolar BECs were reported. In the future, we will use
our new numerical methods to simulate the ground states and dynamics
of dipolar BEC with experimental relevant setups and extend our
methods for rotating dipolar BECs.

\bigskip

\begin{center}
 {\bf Acknowledgements}
\end{center}

This work was supported in part by the Academic Research Fund of
Ministry of Education of Singapore grant R-146-000-120-112 (W.B.,
Y.C. and H.W.) 
and the National Natural Science Foundation of China grant 10901134
(H.W.). We acknowledge very stimulating and helpful discussions with
Professor Peter A. Markowich on the topic.  This work was partially
done while the authors were visiting the Institute for Mathematical
Sciences, National University of Singapore, in 2009.

\bigskip

\begin{center}
 {\bf Appendix  \qquad Proof of the equality (\ref{decop1})}
\end{center}

\renewcommand{\theequation}{\Alph{section}.\arabic{equation}}
\setcounter{equation}{0} \setcounter{section}{1}

Let \be \phi(\bx)=\frac{1}{r^3}\left(1-\frac{3(\bx\cdot {\bf
n})^2}{r^2}\right), \qquad r=|\bx|, \qquad \bx\in{\Bbb R}^3.\ee For
any $\bn\in{\Bbb R}^3$ satisfies $|\bn|=1$, in order to prove
(\ref{decop1}) holds in the distribution sense, it is equivalent to
prove the following:
  \be \label{tt87}
\int_{{\Bbb R}^3} \phi(\bx) f(\bx) d\bx =-\frac{4\pi}{3}f({\bf 0}) -
\int_{{\Bbb R}^3}f(\bx)\; \p_{\bn\bn} \left(\fl{1}{r}\right)  d\bx,
\qquad \forall f(\bx) \in C_0^\infty ({\Bbb R}^3). \label{theorem1}
\ee For any fixed $\vep>0$, let $B_\varepsilon=\{\bx\in\Bbb R^3\ |\
|\bx|<\varepsilon\}$ and $B_\varepsilon^c=\{\bx\in\Bbb R^3 \ |\
|\bx|\ge \varepsilon\}$. It is straightforward to check that \be
\label{phi1234}\phi(\bx)=-\p_{\bn\bn}\left(\frac 1r\right), \qquad
0\ne \bx\in {\Bbb R}^3.\ee Using integration by parts and noticing
(\ref{phi1234}), we get \bea\label{vkder} \int_{B_\varepsilon^c }
\phi(\bx) f(\bx) d\bx&=&- \int_{B_\varepsilon^c}f(\bx)\; \p_{\bn\bn}
\left(\fl{1}{r}\right)  \; d\bx \nn \\
&=&\int_{B_\varepsilon^c}\p_\bn \left(\fl{1}{r}\right)\; \p_\bn
(f(\bx))\; d\bx +\int_{\p B_\varepsilon}f(\bx)\;\frac{{\bf
n}\cdot\bx}{r}\;\p_\bn
\left(\fl{1}{r}\right)\,  dS\nn\\
&=&-\int_{B_\vep^c}\frac{1}{r}\;\p_{\bn\bn}(f(\bx))\;d\bx+I^\vep_1+I^\vep_2,
    \eea
where \be \label{I1I2}I^\vep_1:=\int_{\p
B_\varepsilon}f(\bx)\;\frac{{\bf n}\cdot\bx}{r}\;\p_\bn
\left(\fl{1}{r}\right)\, dS, \qquad I^\vep_2:=-\int_{\p
B_\varepsilon}\frac{{\bf n}\cdot\bx}{r^2}\;\p_\bn
\left(f(\bx)\right)\,dS. \ee From (\ref{I1I2}), changing of
variables, we get
\begin{eqnarray} \label{I145}I^\vep_1&=&-\int_{\p
B_\vep}\frac{({\bf n}\cdot \bx)^2}{r^4}f(\bx)\,dS=-\int_{\p
B_1}\frac{({\bf n}\cdot
\bx)^2}{\vep^2}f(\vep\bx)\,\vep^2dS\nn\\
&=&-\int_{\p B_1}({\bf n}\cdot \bx)^2f({\bf 0})\,dS-\int_{\p
B_1}({\bf n}\cdot \bx)^2\left[f(\vep\bx)-f({\bf 0})\right]\,dS.
\end{eqnarray} Choosing $0\ne\bn_1\in{\Bbb R}^3$ and $0\ne \bn_2\in{\Bbb R}^3$
such that $\{{{\bf{n}_1},\,{\bf{n}_2}\,\bf{n}}\}$ forms an
orthornormal basis of ${\Bbb R}^3$, by symmetry, we obtain
\bea\label{etd11} &&A:=\int_{\p B_1}({\bf n}\cdot
\bx)^2\,dS=\frac{1}{3}\int_{\p
B_1}\left[({\bf{n}}\cdot\bx)^2+({\bf{n}_1}\cdot\bx)^2+
({\bf{n}_2}\cdot\bx)^2\right]\,dS\nn\\
&&\quad =\frac{1}{3}\int_{\p B_1} |\bx|^2dS=\frac{1}{3}\int_{\p B_1}dS=\frac{4\pi}{3},\\
\label{etd12} &&\left|\int_{\p B_1}({\bf n}\cdot
\bx)^2\left(f(\vep\bx)-f({\bf 0})\right)\,dS\right|=\left|\int_{\p
B_1}({\bf
n}\cdot \bx)^2\vep\; \left[\bx\cdot \nabla f(\tht\vep\bx)\right]\,dS\right|\nn\\
&&\quad \leq\vep \,\|\nabla f\|_{L^\infty(B_\vep)} \int_{\p
B_1}\,dS\leq 4\pi\vep\,\|\nabla f\|_{L^\infty(B_\vep)},
\end{eqnarray}
where $0\le \tht\le 1$. Plugging (\ref{etd11}) and (\ref{etd12})
into (\ref{I145}), we have \be \label{I167} I^\vep_1\to
-\frac{4\pi}{3}f({\bf 0}), \qquad \vep \to 0^+. \ee Similarly, for
$\vep\to0^+$, we get \bea\label{I267} &&|I^\vep_2|\leq \|\nabla
f\|_{L^\infty(B_\vep)}\int_{\p B_\vep} \frac{1}{\vep}\,dS=
4\pi\vep\, \|\nabla
f\|_{L^\infty(B_\vep)}\to0, \\
 \label{chat34} &&\left|\int_{B_\vep}
\frac{1}{r}\;\p_{\bn\bn}(f(\bx))\,d\bx\right|\le
\|D^2f\|_{L^\infty(B_\vep)}\, \int_{B_\vep} \frac{1}{r}\,d\bx \leq
2\pi\vep^2\,\|D^2f\|_{L^\infty(B_\vep)} \to 0. \qquad \qquad \eea
Combining (\ref{I167}), (\ref{I267}) and (\ref{chat34}), taking
$\vep\to0^+$ in (\ref{vkder}), we obtain
  \be \label {tt98}\int_{{\Bbb R}^3}
\phi(\bx) f(\bx) d\bx =-\fl{4\pi}{3} f({\bf 0}) - \int_{{\Bbb R}^3}
\fl{1}{r}\;\p_{\bn\bn} (f(\bx)) \,d\bx, \qquad \forall f(\bx) \in
C_0^\infty ({\Bbb R}^3). \label{theorem3} \ee Thus (\ref{tt87})
follows from (\ref{tt98}) and the definition of the derivative in
the distribution sense, i.e. \be \int_{{\Bbb R}^3} f(\bx)\;
\p_{\bn\bn} \left(\fl{1}{r}\right)
      d\bx = \int_{{\Bbb R}^3}\fl{1}{r}\;\p_{\bn\bn}
      (f(\bx))\,d\bx, \qquad \forall f(\bx) \in
C_0^\infty ({\Bbb R}^3),\ee and the equality (\ref{decop1}) is
proven. $\Box$


\begin{thebibliography}{s99}

\bibitem{Abad}
M. Abad, M. Guilleumas, R. Mayol and M. Pi, Vortices in
Bose-Einstein condensates with dominant dipolar interactions, Phys.
Rev. A, 79 (2009), article  063622.

\bibitem{Ant}
P. Antonelli and C. Sparber, Existence of solitary waves in dipolar
quantum gases, preprint.

\bibitem{Bao3}W. Bao,
 Ground states and dynamics of multi-component Bose-Einstein condensates,
Multiscale Model. Simul., 2 (2004),  pp. 210-236.

\bibitem{Bao1} W. Bao and Q. Du,
 Computing the ground state solution of Bose-Einstein condensates by
 a normalized gradient flow, SIAM J. Sci. Comput., 25 (2004), pp. 1674-1697.

\bibitem{Bao_Jaksch_Markowich}
W. Bao, D. Jaksch and P. A. Markowich, Numerical solution of the
Gross-Pitaevskii equation
 for Bose-Einstein condensation,
J. Comput. Phys., 187 (2003), pp. 318-342.

\bibitem{Bao6}
W. Bao, I-L. Chern and F. Y. Lim, Efficient and spectrally accurate
numerical methods for computing ground and first excited states in
Bose-Einstein condensates, J. Comput. Phys.,  219 (2006),  pp.
836-854.

\bibitem{Bao5}
W. Bao and W. Tang, Ground state solution of Bose-Einstein
condensate by directly minimizing the energy functional, J. Comput.
Phys., 187 (2003), pp. 230-254.

\bibitem{Bao2}
 W. Bao, H.   Wang and P. A. Markowich, Ground state, symmetric and central
 vortex state in rotating Bose-Einstein condensate, Comm. Math. Sci., 3 (2005),
 pp. 57-88.


\bibitem{Bao8}
W. Bao and Y. Zhang, Dynamics of the ground state and central vortex
states in Bose-Einstein condensation, Math. Models Meth. Appl. Sci.,
 15 (2005),  pp. 1863-1896.


\bibitem{Blakie}
P. B. Blakie, C. Ticknor, A. S. Bradley, A. M. Martin, M. J. Davis
and Y. Kawaguchi, Numerical method for evolving the dipolar
projected Gross-Pitaevskii equation, Phys. Rev. E, 80 (2009),
aritcle 016703.

\bibitem{Ca1}M. Caliari, A. Ostermann, S. Rainer and M. Thalhammer,
 A minimisation approach for computing the ground state of
Gross-Pitaevskii systems, J. Comput. Phys., 228 (2009), pp. 349-360.




\bibitem{Carles}
R. Carles, P. A Markowich and C. Sparber, On the Gross-Pitaevskii
equation for trapped dipolar quantum gases, Nonlinearity, 21 (2008),
pp. 2569-2590.



\bibitem{Cazen}
T. Cazenave, Semilinear Schr\"{o}dinger equations, (Courant Lecture
Notes in Mathematics vol. 10), New York University, Courant
Institute of Mathematical Sciences, AMS,  2003.

\bibitem{Chang1}
 S. M. Chang, W. W. Lin and S. F. Shieh,
Gauss-Seidel-type methods for energy states of a multi-component
Bose-Einstein condensate, J. Comput. Phys., 202 (2005), pp. 367-390.

\bibitem{Tosi}
 M. L. Chiofalo, S. Succi and M. P. Tosi, Ground state of
trapped interacting Bose-Einstein condensates by an explicit
imaginary-time algorithm, Phys. Rev. E,  62 (2000), pp. 7438-7444.



\bibitem{Eberlein}
C. Eberlein, S. Giovanazzi, and D. H. J. O$'$ Dell, Exact solution
of the Thomas-Fermi equation for a trapped Bose-Einstein condensate
with dipole-dipole interactions, Phys. Rev. A, 71 (2005), article
033618.

\bibitem{Ellio}
M. S. Ellio, J. J. Valentini, and D.W. Chandler,  Subkelvin cooling
NO molecules via "billiard-like" collisions with argon, Science, 302
(2003), pp. 1940-1943.

\bibitem{Giovanazzi}
S. Giovanazzi, P. Pedri, L. Santos, A. Griesmaier, M. Fattori, T.
Koch, J. Stuhler and T. Pfau, Expansion dynamics of a dipolar
Bose-Einstein condensate, Phys. Rev. A, 74 (2006), article  013621.

\bibitem{Glaum}
K. Glaum and A. Pelster,  Bose-Einstein condensation temperature of
dipolar gas in anisotropic harmonic trap, Phys. Rev. A, 76 (2007),
article 023604.

\bibitem{Goral}
K. Go'ral, K. Rzayewski and T. Pfau, Bose-Einstein condensation with
magnetic dipole-dipole forces, Phys. Rev. A, 61 (2000), 051601(R).

\bibitem{Goral1}
K. Go'ral and L. Santos, Ground state and elementary excitations of
single and binary Bose-Einstein condensates of trapped dipolar
gases, Phys. Rev. A, 66 (2002), article 023613.

\bibitem{Griesmaier}
A. Griesmaier, J. Werner, S. Hensler, J. Stuhler and T. Pfau,
Bose-Einstein condensation of Chromium, Phys. Rev. Lett., 94 (2005),
article 160401.


\bibitem{Jiang}
T. F. Jiang and W. C. Su, Ground state of the dipolar Bose-Einstein
condensate, Phys. Rev. A, 74 (2006), article 063602.


\bibitem{Klawunn}
M. Klawunn, R. Nath, P. Pedri and L. Santos, Transverse instability
of straight vortex lines in dipolar Bose-Einstein condensates, Phys.
Rev. Lett., 100 (2008), article 240403.





\bibitem{Lahaye1}
T. Lahaye, J. Metz, B. Fr\"{o}hlich, T. Koch, M. Meister, A.
Griesmaier, T. Pfau, H. Saito, Y. Kawaguchi and M. Ueda, D-wave
collapse and explosion of a dipolar Bose-Einstein condensate, Phys.
Rev. Lett., 101 (2008), article 080401.

\bibitem{Lie}
 E. H. Lieb, R. Seiringer and J. Yngvason, Bosons in a
trap: a rigorous derivation of the Gross-Pitaevskii energy
functional, Phy. Rev. A,  61 (2000), article 043602.

\bibitem{Nath}
R. Nath, P. Pedri and L. Santos, Soliton-soliton scattering in
dipolar Bose-Einstein condensates, Phys. Rev. A, 76 (2007),
 article 013606.


\bibitem{Odell0}
D. H. J. O$'$Dell, S. Giovanazzi and C. Eberlein, Exact
hydrodynamics of a trapped dipolar Bose-Einstein condensate, Phys.
Rev. Lett., 92 (2004), article 250401.

\bibitem{ODell}
D. H. O$'$Dell and C. Eberlein,
 Vortex in a trapped Bose-Einstein
condensate with dipole-dipole interactions, Phys. Rev. A, 75 (2007),
article 013604.

\bibitem{Parker}
N. G. Parker, C. Ticknor, A. M. Martin and D. H. J. O'Dell1,
Structure formation during the collapse of a dipolar atomic
Bose-Einstein condensate, Phys. Rev. A, 79 (2009), article 013617.

\bibitem{Pitaevskii}
L. P. Pitaevskii and S. Stringari, Bose-Einstein Condensation,
Oxford University, New York, 2003.

\bibitem{Pedri}
P. Pedri and L. Santos, Two-dimensional bright solitons in dipolar
Bose-Einstein condensates, Phys. Rev. Lett., 95 (2005), article
200404.

\bibitem{Recati}
A. Recati, I. Carusotto, C. Lobo and S. Stringari, Dipole
polarizability of a trapped superfluid Fermi gas, Phys. Rev. Lett.,
97 (2006), article 190403.

\bibitem{Ronen}
S. Ronen, D. C. E. Bortolotti and J. L. Bohn, Bogoliubov modes of a
dipolar condensate in a cylindrical trap, Phys. Rev. A, 74 (2006),
article 013623.

\bibitem{Sage}
J. M. Sage, S. Sainis, T. Bergeman and D. DeMille, Optical
production of ultracold polar molecules, Phys. Rev. Lett., 94
(2005), article 203001.

\bibitem{Santos} L. Santos, G. Shlyapnikov, P. Zoller and
M. Lewenstein, Bose-Einstein condesation in trapped dipolar gases,
Phys. Rev. Lett., 85 (2000), pp. 1791-1797.

\bibitem{Schnerder}
 B. I. Schneider and D. L. Feder, Numerical approach to
the ground and excited states of a Bose-Einstein condensed gas
confined in a completely anisotropic trap, Phys. Rev. A, 59 (1999),
pp. 2232-2242.


\bibitem{ST}
J. Shen and T. Tang,  Spectral and High-Order Methods with
Applications, Science Press, Beijing, 2006.

\bibitem{str}
 G. Strang,  On the construction and comparison of difference
schemes, SIAM J. Numer. Anal., 5 (1968), pp. 505-517.

\bibitem{Sulem}
C. Sulem and P.-L. Sulem, The nonlinear Schr\"{o}dinger equation,
self-focusing and wave collapse, Springer-Verlag, New York, 1999.

\bibitem{Tick}
C. Ticknor, N. G. Parker, A. Melatos, S. L. Cornish, D. H. J. O'Dell
and A. M. Martin, Collapse times of dipolar Bose-Einstein
condensates, Phys. Rev. A,  78 (2008),    article  061607.



\bibitem{Tikhonenkov}
I. Tikhonenkov, B. A. Malomed and A. Vardi, Anisotropic solitons in
dipolar Bose-Einstein condensates, Phys. Rev. Lett., 100 (2008),
article 090406.

\bibitem{Wang}
D. Wang,  J. Qi, M. F. Stone, O. Nikolayeva, H. Wang, B. Hattaway,
S. D. Gensemer, P. L. Gould, E. E. Eyler and W. C.  Stwalley,
Photoassociative production and trapping of ultracold KRb molecules,
Phys. Rev. Lett.,  93 (2004), article 243005.

\bibitem{Wilson}
R. M. Wilson, S. Ronen and J. L. Bohn, Stability and excitations of
a dipolar Bose-Einstein condensate with a vortex, Phys. Rev. A, 79
(2009), article 013621.

\bibitem{Wilson1}
R. M. Wilson, S. Ronen, J. L. Bohn, and H. Pu, Manifestations of the
roton mode in dipolar Bose-Einstein condensates, Phys. Rev. Lett.,
100 (2008), article 245302.

\bibitem{Xiong}
B. Xiong, J. Gong, H. Pu, W. Bao and B. Li, Symmetry breaking and
self-trapping of a dipolar Bose-Einstein condensate in a double-well
potential, Phys. Rev. A, 79 (2009), article 013626.

\bibitem{Yi0}
S. Yi and H. Pu, Vortex structures in dipolar condensates, Phys.
Rev. A, 73 (2006), article 061602(R).

\bibitem{Yi}
S. Yi and L. You, Trapped atomic condensates with anisotropic
interactions, Phys. Rev. A, 61 (2000),  article 041604(R).

\bibitem{Yi2}
S. Yi and L. You, Trapped condensates of atoms with dipole
interactions, Phys. Rev. A, 63 (2001), article 053607.

\bibitem{Yi1}
S. Yi and L. You, Expansion of a dipolar condensate, Phys. Rev. A,
67 (2003), article 045601.

\bibitem{Yi3}
S. Yi and L You, Calibrating dipolar interaction in an atomic
condensate, Phys. Rev. Lett., 92 (2004), article 193201.

\bibitem{Zhang}
J. Zhang and H. Zhai, Vortex lattice in planar Bose-Einstein
condendsates with dipolar interactions, preprint.

\end{thebibliography}
\end{document}